\documentclass{aa}
\usepackage{natbib}
\usepackage[version=4]{mhchem}
\usepackage[normalem]{ulem}
\usepackage{longtable}
\usepackage[dvipsnames]{xcolor}
\usepackage{caption}
\usepackage{placeins}
\usepackage{float}
\usepackage{textcomp}
\bibpunct{(}{)}{;}{a}{}{,} 
\usepackage{xcolor}
\usepackage{graphicx}
\usepackage{txfonts}
\usepackage{amsmath}    
\usepackage{amssymb}    
\usepackage[]{hyperref}
\begin{document} 

   \title{Characterising the atmosphere of 55 Cancri e}
    \subtitle{1D forward model grid for current and future JWST observations}

   \author{M. Zilinskas\inst{1}\thanks{Currently employed by the Jet Propulsion Laboratory, California Institute of Technology.} \and
          C.P.A. van Buchem\inst{2}\and
          S. Zieba\inst{3} \and
          Y. Miguel\inst{1,2} \and
          E. Sandford\inst{2} \and
          R. Hu\inst{4} \and
          J. A. Patel\inst{5}
          A. Bello-Arufe\inst{4}\and
          L.J. Janssen\inst{2} \and
          S.-M. Tsai\inst{6} \and
          D. Dragomir\inst{7} \and
          Michael Zhang\inst{8}
          }

   \institute{SRON Netherlands Institute for Space Research , Niels Bohrweg 4, 2333 CA Leiden, the Netherlands
   \and
    Leiden Observatory, Leiden University, Niels Bohrweg 2, 2333CA Leiden, the Netherlands
    \and
    Center for Astrophysics, Harvard \& Smithsonian, 60 Garden Street, Cambridge, MA 02138, USA
    \and
    Jet Propulsion Laboratory, California Institute of Technology, Pasadena, CA, USA.
    \and
    Department of Astronomy, Stockholm University, AlbaNova University Center, 10691 Stockholm, Sweden
    \and
    Department of Earth Sciences, University of California, Riverside, CA 92521, USA
    \and
    Department of Physics and Astronomy, University of New Mexico, 210 Yale Boulevard, Albuquerque, NM 87131, USA
    \and
    Department of Astronomy and Astrophysics, University of Chicago, Chicago, IL, USA.
    \\
            \\
              \email{mantas.zilinskas@jpl.nasa.gov}
             }

   \date{Received June XX, 2024; accepted July XX, 2025}

 
  \abstract
  {Recent JWST observations with NIRCam and MIRI of the ultra-short-period super-Earth 55 Cancri e indicate a possible volatile atmosphere surrounding the planet. Previous analysis of the NIRCam spectra suggested potential absorption features from \ce{CO2} or \ce{CO} and significant sub-weekly variability. The MIRI low-resolution spectrum does not contain substantial features but was found to be consistent with effective heat redistribution models. For this study, we computed a grid of over 25000 self-consistent 1D forward models incorporating H-N-O-C-S-P-Si-Ti equilibrium chemistry and assessed plausible atmospheric compositions based on the current JWST data. Despite exhaustive analysis, the composition and properties of the atmosphere remain elusive. While our results statistically favour a global, hydrogen-free, nitrogen-dominated atmosphere enriched in \ce{PO} and \ce{CO2}, various alternative compositions, including \ce{H2O}-,\ce{CO}-, \ce{PH3}-, or Si-bearing, remain viable explanations. Unconstrained heat redistribution efficiency and absolute NIRCam flux are among the largest sources of uncertainty in our analysis. We also find that the heat redistribution factor and surface pressure are highly degenerate with atmospheric composition, and that these parameters cannot be independently constrained using current JWST observations. Furthermore, we show that the observed variability may arise from dynamic interactions between the atmosphere and an underlying magma ocean, driving rapid shifts in atmospheric chemistry and thermal emission. Our results highlight the importance of using self-consistent forward models when analysing novel JWST spectra with limited signal-to-noise ratios -- such as those of 55 Cancri e -- as it allows for a more comprehensive evaluation of potential atmospheric scenarios while also being less sensitive to subtle spectral differences than retrievals. Future JWST observations, particularly at longer wavelengths with MIRI imaging mode to obtain broadband photometry, could help mitigate compositional degeneracies and provide further insight into variability. Constraining the heat redistribution value through phase curve measurements would also significantly reduce degeneracies. For a more complete characterisation of this iconic super-Earth, high-precision spectra are essential.}


   \maketitle


\section{Introduction}
During the first general observers (GO) cycle, the James Webb Space Telescope (JWST) captured thermal emission spectra of the ultra-short-period (USP) super-Earth 55 Cancri e. It was observed using NIRCam and MIRI low-resolution modes as part of \citet{Hu_2024} (GO 1952) and again using NIRCam for a separate programme by \citet{Patel_2024} (GO 2084). 

The NIRCam spectrum from \citet{Hu_2024}, obtained from a single visit, contains an absorption feature at 4 -- 5 \textmu m, which may be due to the presence of \ce{CO2}, \ce{CO}, or other mid-infrared (MIR) absorbers. However, the observation was affected by substantial correlated noise, providing only relative eclipse depths. Wavelengths probed by MIRI show a nearly featureless low-flux continuum, which is consistent with effective heat redistribution, indicative of a global atmosphere \citep{Hammond_2017}. However, due to the lack of strong evidence for known absorbers in the MIRI spectrum, combined with the unconstrained absolute flux of NIRCam, initial analysis resulted in significant degeneracies among the fitted models \citep{Hu_2024}.

The NIRCam results from \citet{Patel_2024} are more ambiguous, showing significant non-correlated flux variability between each of their four visits. 55 Cancri e has long been known to exhibit a high degree of variability \citep{Dragomir_2014,Demory_2016b,Sulis_2019,Valdes_2022}, making it notoriously challenging to characterise. Various explanations have been proposed, including plumes and dust tori created by active volcanism \citep{Demory_2016b,Morris_2021,Meier_2023}, transient atmospheres \citep{Heng_2023}, star-planet interaction \citep{Bourrier_2018,Folsom_2020}, and feedback mechanisms between melt vaporisation and condensation of silicates \citep{Loftus_2024}. \citet{Patel_2024} investigated whether the variability could be caused by asynchronous rotation of the planet, but found no evidence for it. The rapidly changing emission flux observed by \citet{Patel_2024} could be indicative of stochastic outgassing of \ce{CO}/\ce{CO2}. Their measured dayside temperature is also lower than what is expected from a bare-rock scenario, supporting the presence of a heat redistributing atmosphere.

Although recent JWST observations of 55 Cancri e provide some of the best evidence to date of an atmosphere surrounding a rocky exoplanet, it still remains one of the most enigmatic worlds discovered. It is undetermined whether the observed 4.3 \textmu m feature by \citet{Hu_2024} is caused by molecular absorption or if an atmosphere is the definitive explanation for the low observed MIRI flux. Furthermore, the origins of the extreme emission variability remain poorly understood. 

The forward model analysis done in \citet{Hu_2024} accounted for the equilibrium chemistry of varied C-H-O-N-S-P compositions. Their results heavily suggest that the atmosphere is \ce{CO}- or \ce{CO2}-rich with a sub-unity C/O ratio. Models that include consistent C-H-N-S chemistry in equilibrium with the underlying magma, which account for varying redox conditions, show similar results, with abundant \ce{CO} and \ce{CO2} gases being the most plausible explanation. However, the spectrum is best fit when additional absorbers, such as \ce{H2O}, \ce{SO2}, or \ce{PH3}, are present. The resulting $\chi^2$ distribution also showed that \ce{PH3} alone at a $10^{-4}$ volume mixing ratio can fit the observed NIRCam feature, removing the requirement for abundant carbon and oxygen in the atmosphere. In addition, their analysis of melt vaporisation models indicate that the observed low emission flux of the planet cannot be reproduced with tenuous silicate-rich atmospheres, which are unable to effectively redistribute the heat. Silicate-rich models also lack the observed spectral modulation in the 4 -- 5 \textmu m region. The combination of volatiles and melt vaporisation could not reproduce the observed spectrum. The results do, however, show that the sporadic outgassing of silicates could be a viable explanation for the variability measured by Spitzer in the 4.5 \textmu m band \citep{Demory_2016a}. 

For this study, we built upon the analysis in \citet{Hu_2024} and computed a comprehensive grid of over 25000 self-consistent 1D models that incorporate H-N-O-C-S-P-Si-Ti equilibrium chemistry. We determined the potential atmospheric compositions and individual species that are consistent with the captured JWST emission spectra from \citet{Hu_2024} and \citet{Patel_2024}. Our models take into account all of the currently available opacities and address discrepancies in the known thermochemical data. The grid is uniformly constructed to explore the effect of varying temperature regimes. We also self-consistently examined the possible interaction of a present volatile atmosphere with the underlying magma ocean and determined the effect of surface pressure on observability. Additionally, we estimated MIRI photometric noise and show that future observations could help address the high degree of degeneracy present in current fitted models. The resulting emission spectra cover the full JWST wavelength range (0.6 - 28 \textmu m), making the models particularly useful for future JWST observations.

This paper is organised as follows. In the Methods (Section \ref{sec:methods}), we describe our approach in constructing a chemistry grid and a self-consistent pipeline that uses thermochemical equilibrium chemistry, radiative-transfer thermal structure, melt vaporisation, and simulated JWST precision. In the Results (Section \ref{sec:results_main}), we showcase our main findings, including best-fitting atmospheric compositions, as well as trends in abundances and elemental ratios. We examine how different atmospheric properties and interactions with the underlying melt can affect observability and highlight how future observations may help resolve current degeneracies in models. Finally, in the Discussion (Section \ref{sec:Discussion}), we summarise the modelling implications of our results and outline the approach that should be addressed for future studies of 55 Cancri e. The study is concluded in Section \ref{sec:Conclusions}.

\section{Methods}
\label{sec:methods}

\subsection{Constructing the chemistry grid}
\label{sec:methods_chemistry}
In a similar approach to the forward models presented in \citet{Hu_2024}, we explored a large grid of chemical compositions of volatiles in thermochemical equilibrium. Our models account for the H-C-N-O-S-P-Si-Ti chemistry and our initial grid was constructed to uniformly sweep varying mole fractions of each element pair (H+N, C+O, S+P, and Si+Ti). With the exception of the Si+Ti pair, all mole fractions were allowed to vary from 0 to $1-1^{-10}$. Due to their low volatility, Si+Ti abundance was restricted to a maximum of 0.3. For each set of mole fractions, we also varied the elemental ratios of N/H ($10^{-4}$ -- $10^{3}$), C/O ($10^{-3}$ -- $10$), S/P ($10^{-4}$ -- $10^{2}$) and Si/Ti ($10$ -- $10^{2}$). This approach allowed us to examine a wide range of possible atmospheric compositions that have not been comprehensively studied in prior work. To achieve this, we developed a pipeline that is capable of computing large grids of fully self-consistent models, ensuring that the chemistry adheres to the radiative-convective thermal structure. To maximise flexibility, the main grid of models does not account for the possible atmospheric interaction with the underlying melt. The effects of such interactions were explored only for a few selected cases, as described at the end of this section.

For each model within the grid, we simulated an emission spectrum (see Section \ref{sec:methods_structure}) and evaluated the goodness of fit to the observed 55 Cancri e emission spectra. The metric we used to evaluate the goodness of fit is the $\chi^2$ statistic, which we minimised to identify the best-fitting models. This is equivalent to maximum likelihood estimation with a Gaussian likelihood function.

After evaluating $\chi^2$ values across the initial grid, we refined the grid by incorporating new compositions near the models with the lowest $\chi^2$ values. New compositions focused on exploring more mole fractions of carbon and oxygen, as well as additional C/O ratios. In total, we computed more than 25000 unique models.

The equilibrium chemistry was solved using the code \texttt{FastChem}\footnote{https://github.com/exoclime/FastChem} \citep{Stock_2018,Stock_2022,Kitzmann_2024}, for which we used both NIST-JANAF \citep{Chase_1998} and Burcat\footnote{http://garfield.chem.elte.hu/Burcat/burcat.html} (NASA9) thermal data. The majority of our models use NIST-JANAF data, where NASA9 polynomials were only used to compare against alternative chemistry (Section \ref{thermo_data}).

To further explore the impact of atmospheric conditions on the spectrum, we also calculated a selection of models that included the outgassing of vaporised melt species. Vaporisation was computed using \texttt{LavAtmos 2.0}\footnote{https://github.com/cvbuchem/LavAtmos} \citep{Buchem_2022,Buchem_2024}, which calculates the melt-vapour equilibrium for a given melt temperature and composition, while accounting for the effect of the overlying volatiles. For simplicity, we took the melt to be representative of the widely adopted bulk silicate Earth (BSE) composition \citep{Schaefer_2009}.

\subsection{Atmospheric structure and emission spectra}
\label{sec:methods_structure}

To compute the thermal structure of the atmosphere, the pipeline uses the radiative-transfer code \texttt{HELIOS}\footnote{https://github.com/exoclime/HELIOS} \citep{Malik_2017,Malik_2019}, for which we have extended the opacity list to include all species relevant to our chemistry. For the majority of the opacities, we used the DACE\footnote{https://dace.unige.ch/} database or the opacity calculator \texttt{HELIOS-K}\footnote{https://github.com/exoclime/HELIOS-K} \citep{Grimm_2015,Grimm_2021}. As in \citet{Grimm_2021,Zilinskas_2022}, the opacities were approximated using a Voigt fitting profile with a wing cutting length of 100 cm$^{-1}$. The complete reference list and descriptions of the opacities and the corresponding line lists used in this study can be found in Table \ref{table:opacities}. 

We included full convective adjustment using the standard diatomic adiabatic coefficient $\kappa = 2/7$. For heat redistribution values we used three distinct scenarios: $f = 1/4$, $f = 1/3$, $f = 2/3$. This covers the possibilities of the atmosphere redistributing heat globally, semi-globally, and dayside-confined. We found that the low observed brightness temperature ($1796 \pm 88 K$) for 55 Cancri e is likely to be incompatible with larger $f$ factors. \citet{Hammond_2017} have used general circulation models (GCMs) for 55 Cancri e to show that for effective heat redistribution the atmosphere must be of high mean molecular weight and requires at least 10 bar in surface pressure. This is further confirmed by the general optical depth scaling law derived from GCMs \citep{Koll_2022}, where for a 10 bar volatile atmosphere the redistribution factor $f$ is approximately $0.4$ \citep{Hu_2024}. In line with this, we set the atmospheric pressure to be 10 bar for the grid. To demonstrate the effect of surface pressure on the emission spectrum, for a selection of best-fitting models, we also computed cases where the surface pressure was allowed to vary from 1 to 100 bar. Similar analysis was done to address the effect of heat redistribution on observability.

For \texttt{HELIOS} calculations, we account for wavelength ranges between 0.1 and 200 \textmu m at a resolution of $\lambda/\Delta\lambda = 2000$. We used the stellar spectrum from \citet{Hu_2024}, which, for wavelengths between 5 and 12 \textmu m  is derived using the MIRI observation. For wavelengths between 0.8 and 5 \textmu m it uses the \citet{Crossfield_2012} spectrum of 55 Cancri. Outside these limits we used a generated PHOENIX \citep{Husser_2013} and ATLAS \citep{Kurucz_1979,Kurucz_1992,Kurucz_1994} models.

We simulated secondary eclipse spectra using \texttt{petitRADTRANS}\footnote{http://gitlab.com/mauricemolli/petitRADTRANS (we used Version 2)} \citep{Molliere_2019,Molliere_2020} at a resolution of $\lambda/\Delta\lambda = 1000$. For \texttt{petitRADTRANS} we used several additional opacities, notably collision-induced opacities, otherwise where possible, the line lists were kept the same as for \texttt{HELIOS}. The differences are highlighted in Table \ref{table:opacities}. All of the spectral models were generated for the full JWST wavelength range of 0.6 -- 28 \textmu m.

\subsection{Simulated JWST precision for MIRI photometry}
\label{sec:methods_jwst}
In addition to the MIRI/LRS mode used by \citet{Hu_2024} to observe 55 Cancri e spectroscopically, MIRI also includes an imager that can be combined with filters to obtain broadband photometry of a target. The MIRI imaging mode has been recognised as a powerful tool for characterising rocky exoplanets by measuring their brightness temperatures in these filters \citep{Greene_2023,Zieba_2023,DiamondLowe_2023}. This capability is also planned to be used in the 500-hour JWST Director’s Discretionary Time (DDT) programme to search for atmospheres around rocky M-dwarf exoplanets \citep{Redfield_2024}.
For 55 Cancri e, the F1500W (15 \textmu m) filter reaches saturation around just four groups per integration -- below the recommended minimum of five -- when used with the smallest subarray (SUB64). However, photometry remains feasible using the F1800W (18 \textmu m), F2100W (21 \textmu m), and F2500W (25 \textmu m) filters. Using the JWST Exposure Time Calculator \citep{Pontoppidan_2016}, we estimated noise levels for these observations, assuming 7, 12, and 42 groups per integration, respectively, while employing a larger subarray size (SUB128) for more robust background subtraction. 
55 Cancri e has yet to be observed at such long wavelengths, which could provide crucial constraints on its atmospheric composition. In Section \ref{sec:future_observations}, we present an analysis of how future MIRI imager observations could improve our understanding of the planet’s atmosphere.

\section{Results}
\label{sec:results_main}

\subsection{Fitting current JWST data to the model grid}
\label{sec:results_fitting}
We fitted our models with the currently available JWST spectral observations of 55 Cancri e, performing analysis on the NIRCam and MIRI data sets from \citet{Hu_2024} and the NIRCam data from \citet{Patel_2024}. For MIRI/LRS, we used the \texttt{Eureka!} reduction from \citet{Hu_2024}, while for NIRCam, we analysed the \texttt{Eureka!} and the \texttt{SPARTA} reductions from \citet{Hu_2024}, as well as the \texttt{stark} reduction of all five visits from \citet[see their Fig. 3, middle column]{Patel_2024}.

Our results show that, for the current JWST data, there is significant degeneracy between the predicted composition and the properties of the possible atmosphere of 55 Cancri e, as indicated in Figures \ref{fig:F1} and \ref{fig:F2}. This degeneracy does not account for the additional challenge of disentangling atmospheric emission from varying surface albedos \citep{Hu_2012,ParkCoy_2024,Hammond_2025}. The magnitude and the shape of the spectral features of atmospheric emission are dependant on the vertical temperature structure. While a global atmosphere with a certain composition can be a good fit, the data can also be explained by an entirely different composition with a vastly hotter temperature profile that is representative of dayside-confined case. The statistical difference between many such cases is insufficient to put any real constrains on molecular abundances or the type of the atmosphere that the planet possesses. In addition to this, the observed variability in emission and the large discrepancies between different reduction pipelines introduce further uncertainty, complicating the efforts to characterise the atmosphere. 

However, by using grids of atmospheric models, we can attempt to predict the constituents and the properties of the atmosphere, as well as assess whether future observations could allow us to break the overwhelming degeneracies. Due to the large variations in the abundances in our models, many cases may not adhere to the conditions permitted by the outgassing of secondary atmospheres (e.g. \citet{Tian_2024}), which may further constrain the proliferation of plausible scenarios. Nonetheless, given the unknown nature of 55 Cancri e, we adopt a comprehensive approach that considers the full range of possibilities, independently of the planet's evolution history. All of the computed emission spectra, temperature profiles, and associated results can be readily accessed through the GitHub\footnote{https://github.com/zmantas/55-cnc-e} portal.

\subsubsection{Fitting the combined \texttt{Eureka!} spectrum}
\begin{figure*}
    \centering
        \includegraphics[width=1\textwidth]{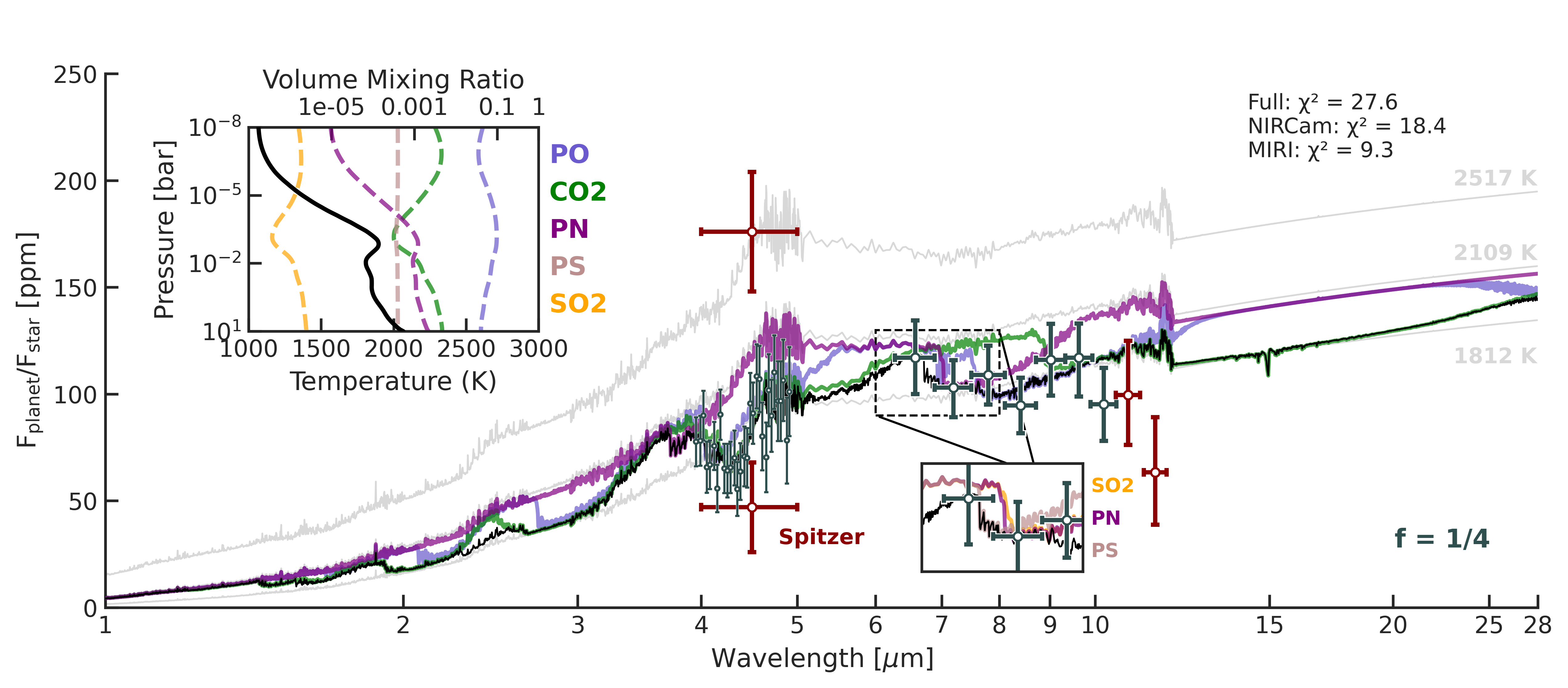}
    \caption{Best-fitting emission spectrum (black) for 55 Cancri e from our results. The coloured spectra show individual contributions of the strongest opacities in the model, \ce{PO} (blue), \ce{CO2} (green), and \ce{PN} (purple). The faint grey spectra, arranged from top to bottom, represent blackbody emission for heat redistribution values of f = 1.0 (substellar temperature), f = 2/3 (dayside-confined), and f = 1/4 (global redistribution), with the equilibrium temperature labelled above each corresponding curve. The spectrum is fitted to the \texttt{Eureka!} reduction of the JWST observations from \citep{Hu_2024} (NIRCam and MIRI/LRS). The NIRCam absolute flux is not constrained, allowing the data to shift. A depth of 79 ppm is used in the figure. The last two points of MIRI beyond 10 \textmu m (dark red) represent the 'shadow region' of the data.  The model assumes heat redistribution coefficient f = 1/4 (global redistribution). The legend in the upper right denotes the calculated $\chi^2$ values for the combined MIRI + NIRCam fit (Full), as well as for the individual NIRCam and MIRI fits to the black spectrum. Overlaid on the spectrum are the two Spitzer (red) observations from \citep{Demory_2016a}. The upper-left inset shows the temperature profile as well as volume mixing ratios of the major absorbers. The lower inset is a zoom in of the 6-8 µm region highlighting the absorption overlap of additional species: \ce{SO2}, \ce{PN}, and \ce{PS}.}
    \label{fig:F1}
\end{figure*}

The \texttt{Eureka!} reduction of the NIRCam and MIRI data points from \citet{Hu_2024} are shown in Figure \ref{fig:F1}. The NIRCam data spans from 4 -- 5 \textmu m, while the MIRI data covers 6 -- 12 \textmu m. Due to saturation of the MIRI detector, there is a discontinuity between the two data sets at 5 -- 6 \textmu m. In addition, the last two MIRI points are affected by the shadowing effect, which, in cases, can change the statistical preference of the model \citep{Hu_2024}. It is also important to note, that due to substantial correlated noise, the absolute flux of the NIRCam spectrum is unconstrained, allowing the data to shift upwards or downwards, introducing further degeneracy between models. 

Figure \ref{fig:F1} showcases the best-fitting emission spectrum model from our grid search. The spectrum denoted in black is a 10 bar global atmosphere with a heat redistribution value of $f = 1/4$ ($T_{eq} = 1968 \, \mathrm{K}$). The coloured spectra indicate the individual contribution of the major absorbers, which are \ce{PO}, \ce{CO2}, and \ce{PN}. The forward models presented in \citet[Extended Data Fig. 8]{Hu_2024}, hinted at a possible influence of phosphorus-containing species on the emission spectrum\footnote{The analysis of Fig. 8 in \citet{Hu_2024} uses the \texttt{SPARTA} reduction, which we analyse later in this section.}. In our extended grid, the phosphorus features are much more preferred. For the displayed case, \ce{CO2} is still the major absorber for NIRCam, but the fit is substantially improved by the present \ce{PO}, which provides additional absorption at 4 \textmu m. At MIRI wavelengths, \ce{PO} has substantial opacity, which aligns well with the observations, even without the presence of \ce{CO2}. For this particular composition, we also find an abundance of \ce{PN}, \ce{PS} and \ce{SO2}, which further improve the fit to the MIRI spectrum, but are degenerate with each other, as indicated by the lower inset panel.

As shown in the upper inset, the temperature structure also exhibits a mild inversion at $10^{-3}$ bar, which is caused by the presence of \ce{PS}. With equilibrium chemistry\footnote{Assuming NIST-JANAF thermochemical data is used for the observed temperatures of 55 Cancri e.}, sulphur in atmospheres containing ample phosphorus and carbon will primarily sequester into \ce{PS}, with \ce{SO2} and \ce{SO} being the lesser sulphur-containing species. This is mainly because the available oxygen is prioritised for the formation of \ce{CO} and \ce{CO2}, as well as \ce{PO} and \ce{PO2}. 

\citet{Schaefer_2012} suggest that the vaporisation of an Earth-like continental crust may result in phosphorus species becoming important atmospheric constituents, with \ce{PO2} and \ce{PO} becoming the main P-carriers at high temperatures expected on lava worlds. \ce{PO} has also been identified as one of the possible dominant phosphorus species in atmospheres of hot Jupiters and warm Neptunes, especially at enhanced metallicities \citep{Lee_2024}. Our models include the opacity of \ce{PO}; however, the line list for \ce{PO2} is not yet available. Its opacity may have a substantial impact on the temperature structure and the emission spectrum of the planet.

The presented model is a significant improvement over blackbody-like cases. The achieved $\chi^2$ for the combined data sets is 27.6. In comparison, the best possible blackbody emission spectrum results in $\chi^2 = 37.0$. The improvement is mostly due to the substantially better NIRCam fit. Excluding MIRI's shadow region reduces the residuals further, making the model even more robust compared to a blackbody spectrum ($\Delta\chi^2 = 11.2$ for model versus blackbody).

\begin{figure}
    \centering
        \includegraphics[width=0.5\textwidth]{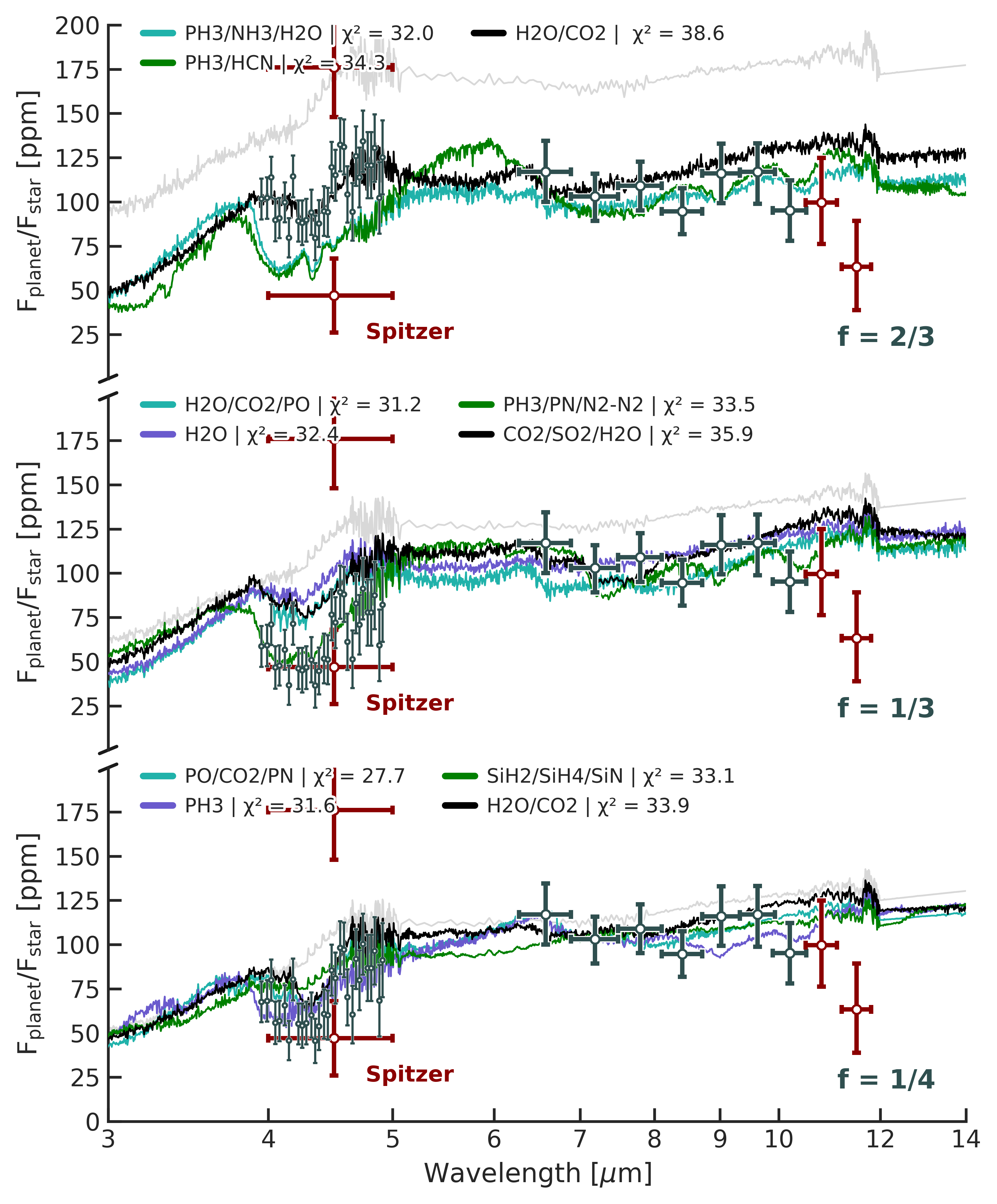}
    \caption{Emission spectra for varying compositions showing degeneracy between heat redistribution and atmospheric composition. Each panel showcases drastically different compositions for each of the heat redistribution $f$ values. Grey curves represent expected blackbody emission for respective $f$ values. The main absorbers contributing to the fit are denoted in the corresponding legends. The calculated $\chi^2$ values are for the \texttt{Eureka!} reduction of the JWST data from \citep{Hu_2024}. The dark red points denote MIRI's shadow region. Note that the absolute flux of NIRCam is allowed to vary. Also shown is the lower flux Spitzer measurement from \citep{Demory_2016a}.}
    \label{fig:F2}
\end{figure}

While a phosphorus-rich atmosphere could explain the JWST data presented in \citet{Hu_2024}, many of the models from our grid have similar $\chi^2$ values\footnote{All the $\chi^2$ tables can be found in https://github.com/zmantas/55-cnc-e}. In Figure \ref{fig:F2}, we show a selection of distinct cases with $\chi^2$ below or near a simple blackbody fit. Each of the panels represents a different temperature regime $f$, ranging from global ($f=1/4$) to dayside-confined redistribution ($f=2/3$). Since MIRI's spectrum is mostly featureless and highly degenerate between broad IR absorption and different $f$ values, it cannot independently constrain the molecular composition of the atmosphere. With the inclusion of NIRCam, we can start decreasing the possibilities, but due to the undefined absolute flux, the number of viable compositions remains large. For well-fitted models, NIRCam's absolute flux offset can reach up to $\approx 110 \, \mathrm{ppm}$, in cases even surpassing the mean flux of MIRI. This creates a proliferation of possible scenarios, ranging from dominant \ce{PH3}, \ce{H2O}, \ce{CO2}, \ce{PO}, to even \ce{SiHx}-rich atmospheres. In some instances, the combined spectrum can even be fitted with a single opacity, such as \ce{H2O} or \ce{PH3}.

To elaborate, in a hot, dayside-confined regime ($f=2/3$, upper panel), the atmosphere can contain strong mid-infrared (MIR) absorbers, such as \ce{H2O}, \ce{HCN}, and even \ce{PH3}, which push the emitting photosphere to $T \approx 1800 \, \mathrm{K}$, matching MIRI's observed flux. While models with a higher equilibrium temperatures are generally harder to fit, they can mimic atmospheres that have cooler, global heat redistribution. However, we find that well-fitting models with hotter temperatures tend to be reducing and often contain abundant \ce{PH3}, as well as other hydrogen-containing gases such as \ce{NH3} or \ce{HCN}, which may be indicative of active photochemistry in a geologically active environment \citep{Rimmer_2019}.

As the global temperature is decreased ($f=1/3$, middle panel), the degeneracy between models increases. More complex and more varied compositions become good fits to the observations. This is primarily due to the lower overall emission flux, which no longer requires any strong continuous absorbers. This allows the atmospheres to be dominated by inert components, such as \ce{N2}, which results in more prolific chemistry. The atmospheres also become more oxidised, making \ce{CO2} a common constituent. Sulphur-rich cases also become viable, as \ce{SO2} opacity can improve the fit to NIRCam's data, while not overwhelming MIRI's portion of the spectrum. In this regime, the observations can also be explained solely with the absorption of silicon compounds, such as \ce{SiH2}, \ce{SiH4}, and \ce{SiN}. Si-rich atmospheres could be a direct indication of a magma ocean that is shrouded with a volatile-filled atmosphere \citep{Zilinskas_2023,Buchem_2024}.

At full heat redistribution (f = 1/4, lower panel), we find the hydrogen-poor and \ce{PO}-rich scenario, which includes abundant nitrogen and \ce{CO2}, producing the best-fitting case discussed earlier in this section. However, just as with other temperature regimes, various compositions can reasonably explain the observations.

\subsubsection{\texttt{Eureka!} analysis for individual instruments}

55 Cancri e is known to show strong emission variability on sub-week timescales \citep{Demory_2016b,Meier_2023,Demory_2023,Patel_2024}. Since the resulting spectrum is different every time the planet is observed, the individual analysis of each instrument data is warranted. The NIRCam and MIRI spectra from \citet{Hu_2024} were taken months apart, which may have probed entirely different atmospheric conditions.

\begin{figure}
    \centering
        \includegraphics[width=0.4\textwidth]{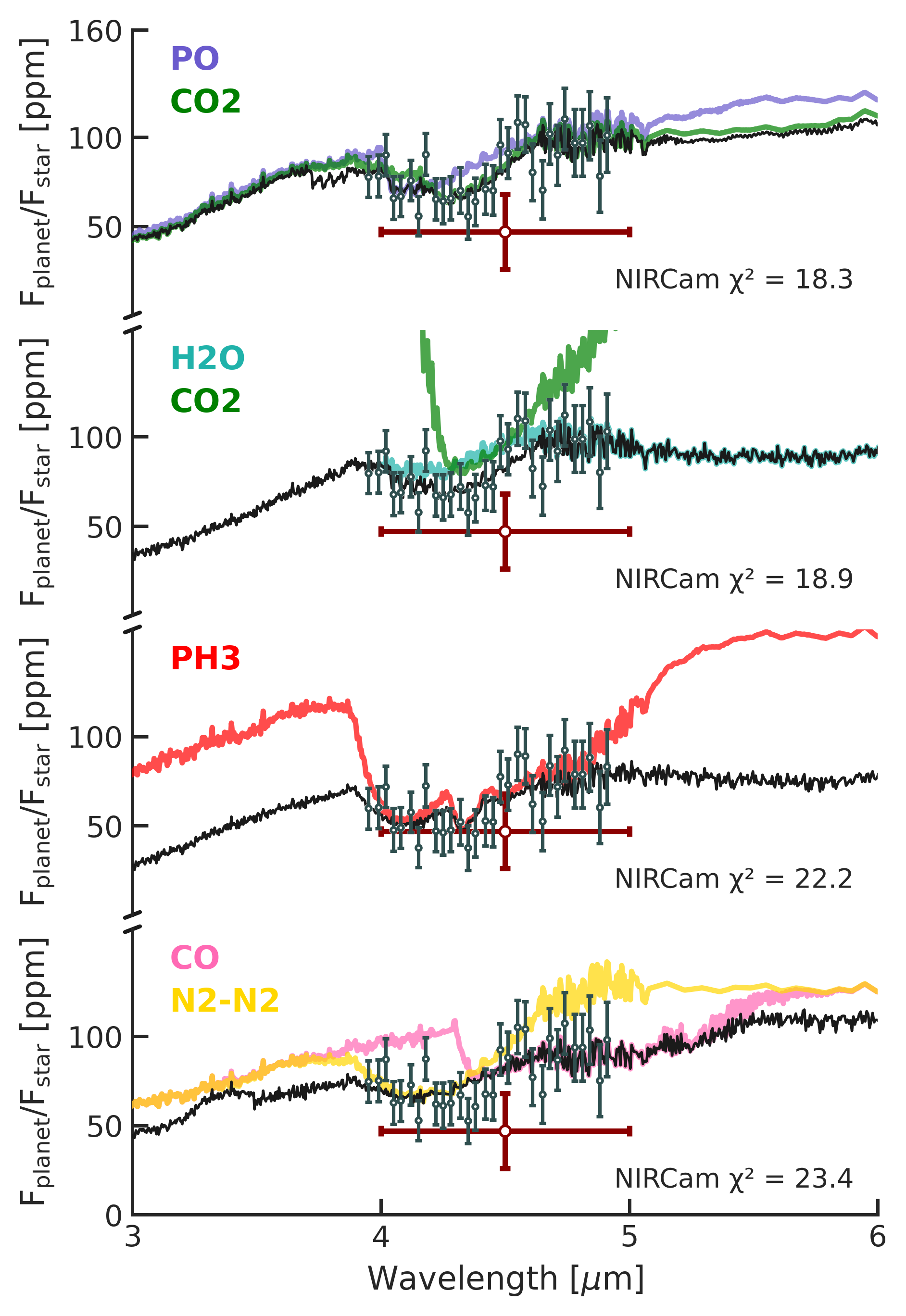}
    \caption{Emission spectra fits only for NIRCam's \texttt{Eureka!} reduction from \cite{Hu_2024}. In each panel, the total spectrum is denoted in black, while the major absorbers are highlighted in colour. The heat redistribution coefficient for the top panel is f = 1/4, the other panels correspond to f = 1/3. The absolute flux of NIRCam is adjusted to achieve the minimum $\chi^2$ values. The best achieved $\chi^2$ value for a blackbody-like spectrum is $27.0$. Also shown in red is the low-flux Spitzer measurement from \citep{Demory_2016a}.}
    \label{fig:F3}
\end{figure}

When fitting the NIRCam spectrum alone, the ambiguity increases. The complexity of the allowed chemistry and the emission dependence on the temperature structure allows for a large number of models to fit the data. Figure \ref{fig:F3} showcases four examples of good fits with better $\chi^2$ values than a simple blackbody model. These few examples effectively highlight the range of possibilities when fitting NIRCam's emission spectrum. Typically, a main absorber, such as \ce{CO2} or \ce{CO} is required, albeit many other options are possible. Since most opacities are not broad enough to cover the entire observed feature, a secondary, complimenting opacity centred around 4 \textmu m, is needed. Secondary absorbers can vary drastically, from \ce{PO}, \ce{H2O}, \ce{SO2}, Si-containing species, to even collision-induced opacities such as \ce{N2-N2}. Inverted temperature structures, where the opacities of \ce{PN} and \ce{PS} dominate, are also possible. Additionally, the spectrum can be fitted with a single opacity, such as \ce{H2O} or, as noted in \citep{Hu_2024}, \ce{PH3} (third panel), regardless of the set temperature regime.

For the \texttt{Eureka!} reduction, the best NIRCam fit matches the showcased combined spectrum, with absorption attributed to the presence of \ce{CO2} and \ce{PO} (top panel). However, \ce{H2O}-dominated atmospheres (second panel), instead of P-dominated ones show almost identical $\chi^2$ values. Constraining the bulk composition or the type of the atmosphere from NIRCam data alone is not yet feasible. As discussed further in this paper, this limitation is inherent to the low precision of the secondary eclipse spectra and is not unique to the \texttt{Eureka!} reduction or this particular JWST visit for 55 Cancri e.

The observed MIRI spectrum by \citet{Hu_2024} closely resembles blackbody emission indicative of $T \approx 1800 \, \mathrm{K}$. This corresponds to an $f$ factor of $\approx0.17$, which is notably lower than the typically assumed global heat redistribution factor of $f=1/4$. A possible explanation for this discrepancy is the presence of a substantial atmosphere with strong MIR opacity in photospheric regions where temperatures fall below the planet's equilibrium temperature. As the observed spectrum does not exhibit any notable features, the data can only be explained via a continuum originating from absorption of a single or more species.

\begin{figure}
    \centering
        \includegraphics[width=0.5\textwidth]{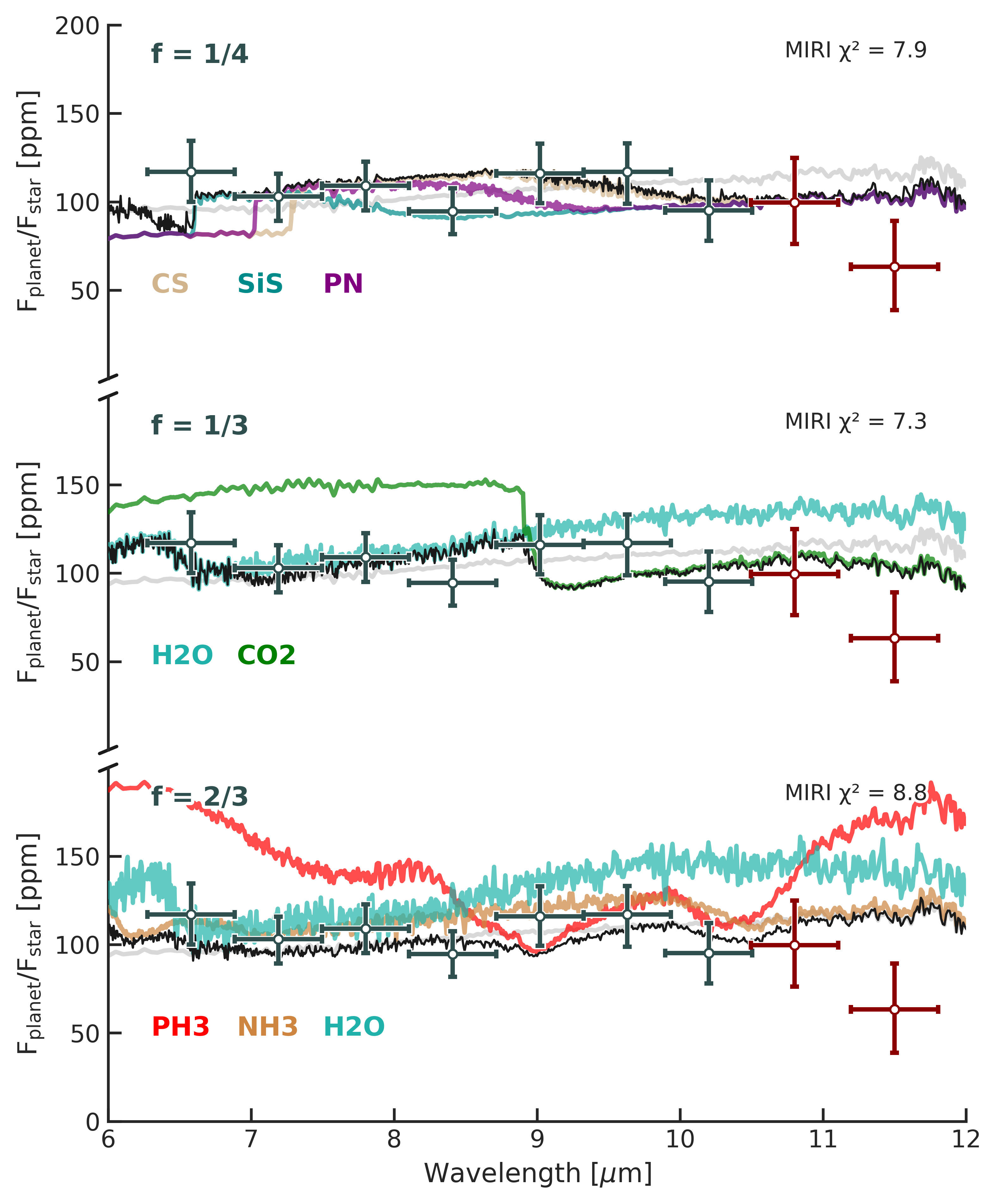}
    \caption{Emission spectra models for MIRI only. The shown data is the \texttt{Eureka!} reduction from \cite{Hu_2024}. Each panel represent a case for the denoted heat redistribution value $f$. The black spectrum is the total model, where the coloured spectra indicate individual contribution of opacities. Shown in grey are the best fitting blackbody spectra corresponding to a $\chi^2 = 9.4$. The two dark red data points are affected by the shadowing effect.}
    \label{fig:F4}
\end{figure}

Figure \ref{fig:F4} illustrates three well-fitting MIRI models (including the shadow region), each corresponding to a different $f$ value. As with NIRCam, a wide variety of different atmospheric compositions can result in good fits. Models rich in \ce{H2O} and \ce{CO2} (as shown in the mid panel) are prime examples of this. The continuum of \ce{H2O} alone can reasonably explain the MIRI spectrum. However, as demonstrated with the model shown in Figure \ref{fig:F1}, \ce{H2O} is not essential, and often not preferred. Many different scenarios fit the observations, including those with inverted temperature profiles (as shown in the upper panel), where opacities are dominated by P- and S-containing species, such as \ce{CS}, \ce{SiS}, and \ce{PN}

At higher temperatures ($f=2/3$) (lower panel), the chemistry and opacities that produce a good fit are typically H-rich. Species such as \ce{PH3}, \ce{NH3}, and \ce{HCN} all have substantial opacities, pushing the photosphere to the required temperatures. Similarly to other temperature regimes, atmospheres with ample \ce{H2O}, \ce{CO2}, and \ce{SO2} also provide good fits.

While we examine specific examples, it is important to emphasise that many other configurations are possible. The MIRI spectrum offers limited constrains the atmospheric composition of 55 Cancri e. Even a blackbody or flat-line model achieves similar $\chi^2$ values, regardless of whether the shadow region is included in the analysis.

\subsubsection{Determining abundances and elemental ratios for the \texttt{Eureka!} reduction}

\begin{figure*}
    \centering
        \includegraphics[width=0.7\textwidth]{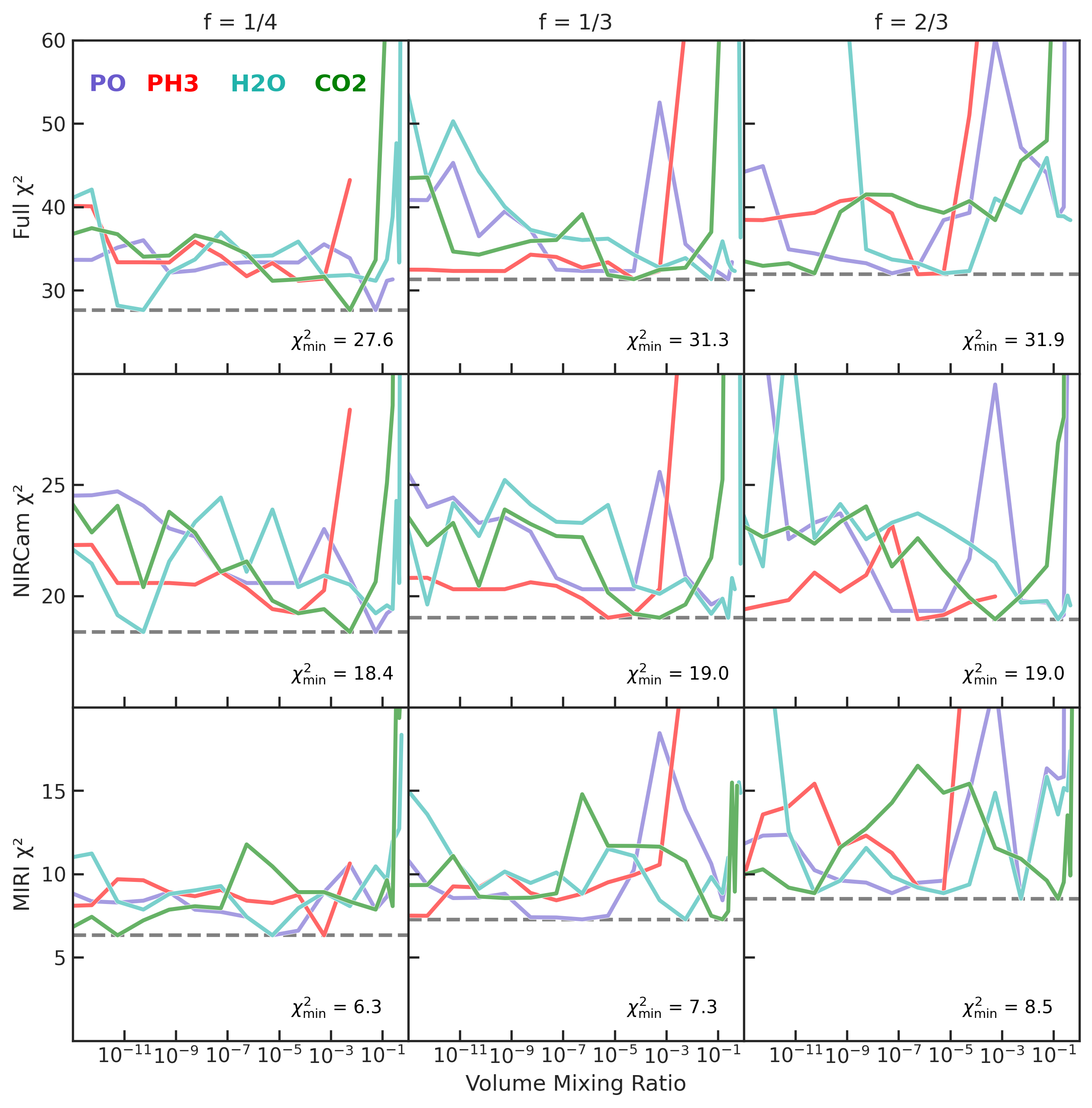}
    \caption{$\chi^2$ dependance on the volume mixing ratio of the indicated species: \ce{PO}, \ce{PH3}, \ce{H2O}, and \ce{CO2}. Shown $\chi^2$ values are the best-achieved for each particular abundance summed over the photospheric region ($P>10^{-5}$ bar). Each column represents a different heat redistribution value $f$. The rows show $\chi^2$ for the combined, or individually denoted spectra of NIRCam and MIRI of the \texttt{Eureka!} reduction from \citet{Hu_2024}. The grey dashed lines represent the overall minimum achieved $\chi^2$ in each regime. }
    \label{fig:F5}
\end{figure*}

It is clear that the spectrum can be reasonably explained using various atmospheric compositions; however, there are certain limits, and perhaps trends in abundances of individual species across the models. In Figure \ref{fig:F5}, we show the $\chi^2$ dependance on the volume mixing ratio of the selected species for the \texttt{Eureka!} reduction. While many different species can be important when fitting the spectra, we selected \ce{PO}, \ce{PH3}, \ce{H2O}, and \ce{CO2}, specifically due to them being some of the largest contributing opacities in most of the good fitting models. 

For the combined spectrum of NIRCam and MIRI (top row), we see that the model with the lowest $\chi^2$ value (27.6), occurring at $f=1/4$, is the same model displayed in Figure \ref{fig:F1}. It has negligible \ce{H2O} abundance and is rich in \ce{PO} and \ce{CO2}. The bulk of this atmosphere is \ce{N2}. While this particular case, within the low temperature regime, stands out from the rest, models with the closest $\chi^2$ values exhibit significant overlap in abundances. For many species, not just for the shown examples, the preferred abundances can vary many orders of magnitude. At best, only an upper limit can be estimated, where the individual opacities become too substantial to fit the spectrum. 

We find similar results for the $f=1/3$ temperature regime. In general, as the temperature is increased the trends become more clear, making the models less degenerate. At the highest temperature regime  ($f=2/3$), due to fewer atmospheric scenarios that can reasonably fit the observed spectrum, abundance limits become even more distinct. An example of this is the drastic $\chi^2$ improvement with decreasing \ce{CO2} abundance. Also, due to the necessity of a strong continuum, the lower limit on \ce{H2O} is constrained at around $> 10^{-9}$ VMR. Unlike in the cooler regimes, the best models here also often involve \ce{PH3}, which has minimised $\chi^2$ at $10^{-7}$ -- $10^{-5}$ volume mixing ratio.

Similar behaviour is reflected in the individual analysis of the instruments (middle and bottom rows); however, NIRCam and MIRI may not necessarily align with each other. NIRCam favours higher \ce{CO2} abundance, whereas the MIRI spectrum is much more ambiguous and uniform, showing no clear preference for this species. As shown in the 3rd column, for $f=2/3$, the combination of the two spectra can drastically shift the trends from individual preferences.

\begin{figure*}
    \centering
        \includegraphics[width=0.8\textwidth]{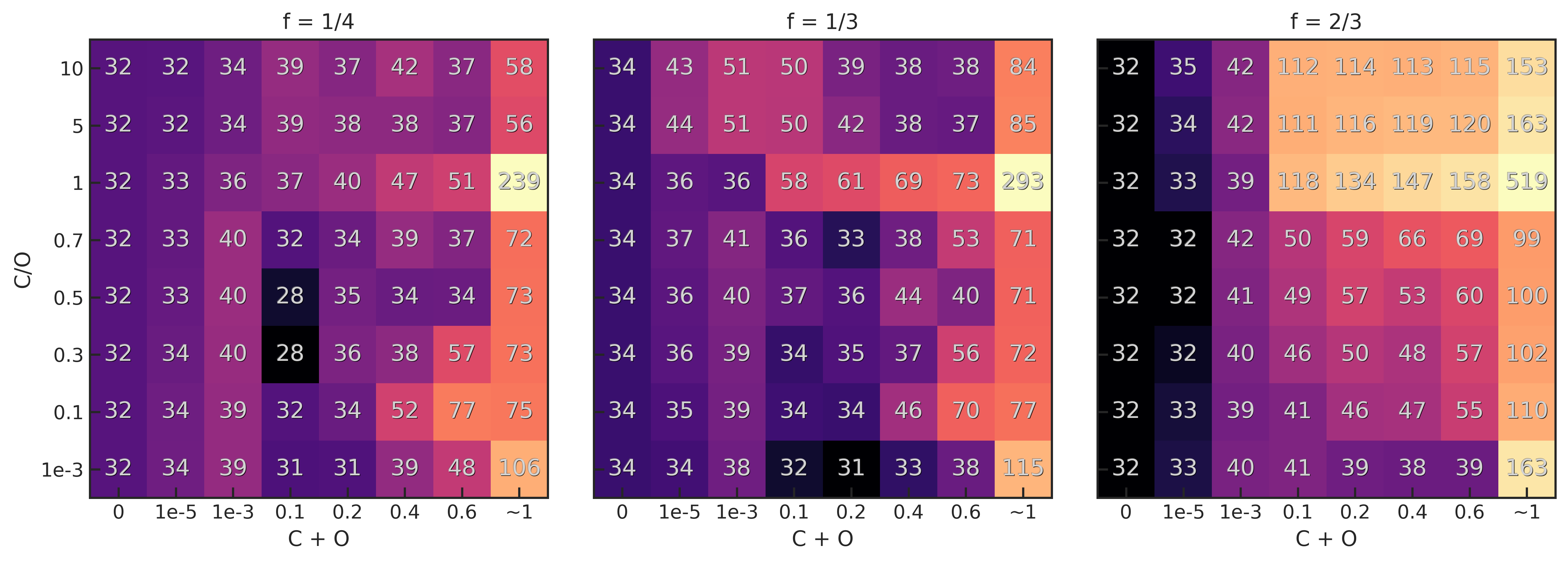}
    \caption{Achieved $\chi^2$ values for different carbon and oxygen mole fractions versus the C/O ratio for \texttt{Eureka!}'s reduction of the combined NIRCam and MIRI spectrum from \citet{Hu_2024}. Each panel represents a different temperature regime, denoted via a $f$ value above it. The colour gradient and the numbers represent the minimal $\chi^2$ values of each case.}
    \label{fig:F6}
\end{figure*}

As noted in the analysis by \citet{Hu_2024,Patel_2024}, as well as our shown results, the atmosphere of 55 Cancri e may be rich in \ce{CO2} or \ce{CO}. Carbon chemistry is dependant on the C/O ratio. Just as in \citet{Hu_2024}, we can map out the statistical preference for the elemental abundances and their respective ratios. In Figure \ref{fig:F6}, we show the achieved $\chi^2$ values against the carbon and oxygen mole fractions and the C/O ratio. Each panel displays the results for a different temperature regime, with the colour gradient representing the minimised $\chi^2$ value, which is also denoted via the individual number labels.

The first immediate conclusion from these results is that the preferred abundance and ratio vary drastically with the set heat redistribution value. At full redistribution ($f=1/4$), the $\chi^2$ is minimised when the C+O mole fraction is 0.1 and the C/O ratio is between 0.3 -- 0.5. A low C/O ratio is required for efficient formation of \ce{CO2}. As also indicated in the previous figure, models with higher $\chi^2$ values are largely degenerate, providing only a vague upper limit on the abundance.

At $f=1/3$, the results are less distinctive. $\chi^2$ values are very similar across the C+O mole fraction range of 0 -- 0.4, indicating that neither \ce{CO2} or \ce{CO} are required for a good fit. Consequently, the C/O ratio is also poorly constrained. Models where \ce{CO2} opacity dominates, with a C/O ratio <= 0.7, fit just as well as \ce{CO}-rich models that have a C/O ratio between 0.7 -- 1.0. Very carbon-rich atmospheres with C/O > 1.0, resulting in abundant hydrocarbons such as \ce{C2H2} can also reasonably explain the observations (a cluster of models at C+O of 0.4 -- 0.6 and C/O of 5 -- 10). However, such compositions are likely to be affected by the condensation of carbon, which will keep the gaseous C/O ratio close to unity \citep{Zilinskas_2020,Hu_2024}.

At dayside-redistributed temperatures ($f=2/3$), we find a clear trend in carbon chemistry, as both \ce{CO} and \ce{CO2} are disfavoured in preference of hydrogen- and phosphorus-rich constituents. Such scenarios were discussed in detail earlier. A similar analysis for each in individual instruments is shown in Appendix \ref{appendixB}.

In summary, for the \texttt{Eureka!} reduction, the models do not strongly constrain the properties or the chemical composition of the atmosphere. The limited precision of the observations and the complex nature of atmospheric chemistry produces significant degeneracies and noise in the $\chi^2$ values, yielding a multitude of likely solutions. Thermal emission is also implicit to the probed temperature structure, making the chemistry highly degenerate with the set heat redistribution factor. In \citet{Hu_2024}, a tenuous, vaporised rock atmosphere -- incapable of efficient heat redistribution -- results in much higher flux than what is observed with MIRI, and is therefore effectively ruled out. Instead, their analysis suggests a substantial, global \ce{CO2}- or \ce{CO}-rich atmosphere, which allows for full heat redistribution ($f=1/4$). For tidally locked planets such as 55 Cancri e, the effectiveness of the heat circulation heavily scales with the mean molecular weight and the surface pressure of the atmosphere \citep{Hammond_2017,Pierrehumbert_2019}.

Our grid of models favours atmospheres rich in \ce{CO2} and \ce{CO}, with \ce{N2} as the background component, along with additional constituents such as \ce{PO}, \ce{PN}, \ce{PS}, \ce{SO2}, and \ce{SO} -- all of which have high mean molecular weights. Moreover, models assuming full heat redistribution are generally better fitting than those with hotter, dayside-confined temperatures.
However, the differences in $\chi^2$ values across various compositions and $f$ factors are insufficient to precisely constrain the atmospheric properties or bulk constituents of the planet. Accounting for variability and analysing individual spectra of each instrument further deepens this degeneracy.

\subsubsection{Alternative reductions and additional visits}
\label{section:alt_reductions}
\begin{figure*}
    \centering
        \includegraphics[width=1\textwidth]{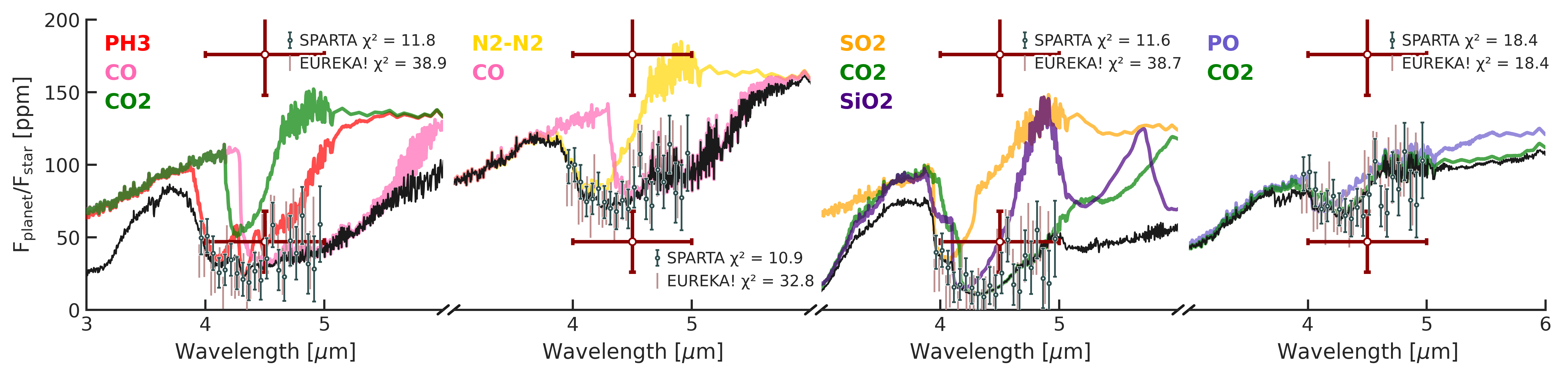}
    \caption{Comparison of the \texttt{SPARTA} and \texttt{Eureka!} reductions for the NIRCam spectrum from \citet{Hu_2024}. The dark grey data points represent \texttt{SPARTA}, while the light brown error bars are for \texttt{Eureka!}. Note that the \texttt{SPARTA} spectrum is lower resolution (21 points, versus 30 for \texttt{Eureka!}), which naturally results in lower $\chi^2$ values. The black spectrum is the total emission, with major absorbers highlighted in colour. Also shown in red is the low-flux Spitzer measurement from \citep{Demory_2016a}.
    }
    \label{fig:F7}
\end{figure*}

In addition to the \texttt{Eureka!} data, we have also analysed the \texttt{SPARTA} reduction from \cite{Hu_2024}, as well as the \texttt{stark} reduction of all five NIRCam visits from \citet{Patel_2024}. In Figure \ref{fig:F7}, we show the comparison between the \texttt{Eureka!} and the \texttt{SPARTA} data. The dark points represent \texttt{SPARTA}, while the faded brown errors are for \texttt{Eureka!}. Although both reductions have strong overlap, the preference for the models is somewhat different. We report the $\chi^2$ value for each model for each reduction. We emphasise that, rigorously, it is only appropriate to compare $\chi^2$ values across different models fit to the same data, and not to compare $\chi^2$ values for the same model, fit to different data sets. Note that the \texttt{SPARTA} reduction is lower resolution, resulting in overall lower $\chi^2$ values.

Just as with \texttt{Eureka!}, many different compositions can reasonably fit the \texttt{SPARTA} reduction. In Figure \ref{fig:F7}, we present a few cases highlighting the potential differences between the two reductions. The rightmost panel shows the best-fitting, hydrogen-free, \ce{PO}- and \ce{CO2}-rich model for the \texttt{Eureka!} reduction. For the \texttt{SPARTA} reduction, the $\chi^2$ value for this composition is substantially higher compared to other displayed cases. Due to the lower flux at 4.0 -- 4.3 \textmu m, \texttt{SPARTA} generally exhibits a stronger preference for hydrogen-rich atmospheres, which leads to better $\chi^2$ values across more varied compositions. Abundant hydrogen also allows for the atmosphere to have plentiful \ce{H2O} and \ce{PH3}. \ce{CO2} can still serve as a primary source of opacity, which in combination with other species, such as \ce{SiO2} or \ce{SO2}, also fit the data well (see third panel). However, with \texttt{SPARTA}, we do see a larger preference towards \ce{CO} being the dominant absorber instead of \ce{CO2} (as illustrated in the first and second panels). 

In the analysis of novel JWST spectra, accounting for multiple data reductions can lead to vastly different interpretations of bulk atmospheric constituents. Even if the features seem similar, small differences can change supporting opacities, which often are the most abundant species in the atmosphere. This is evident in the comparison of the \texttt{SPARTA} and \texttt{Eureka!} reductions. Further analysis of trends in abundances and elemental ratios for \texttt{SPARTA} can be found in Appendix \ref{appendixB}.

\begin{figure*}
    \centering
        \includegraphics[width=1\textwidth]{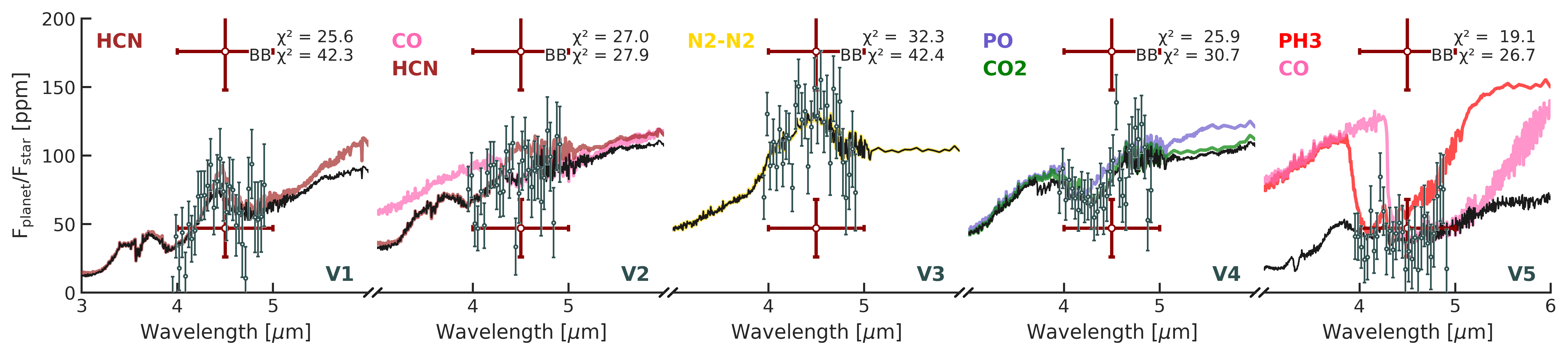}
    \caption{Best-fitting emission spectra models for the \texttt{stark} reduction of all five NIRCam visits from \citet{Patel_2024}. The visits are ordered from left to right and labelled with V1-V5. Visit 4 is the \texttt{stark} reduction of the same data presented in \citet{Hu_2024}. The black spectrum is the total emission, with major absorbers highlighted in colour. The dark red points represent the low-flux Spitzer measurement from \citep{Demory_2016a}. The indicated BB $\chi^2$ values are for the best fitting blackbody model for each visit. The fitted models assume full heat redistribution $f=1/4$.}
    \label{fig:F8}
\end{figure*}

Returning to the topic of variability, we also fit the five NIRCam visits presented in \citet[see their Fig. 3, middle column]{Patel_2024}. Although four of the visits were taken within a week, each spectra are substantially different, showing very distinct preferences for the composition. Figure \ref{fig:F8} displays the \texttt{stark} reduction of their observations, along with our best-fitting models. As with the \citet{Hu_2024} NIRCam data, the absolute flux is unconstrained and allowed to vary freely. The resolution of \texttt{Eureka!} and \texttt{stark} reductions is identical. Note that the fourth visit is the same NIRCam data as in \citet{Hu_2024}, for which we analysed the \texttt{Eureka!} and \texttt{SPARTA} reductions in previous sections. Although the \texttt{stark} reduction of this visit closely resembled \texttt{Eureka!}, we include it here for completeness. All fitted models here assume global heat redistribution ($f=1/4$). While not shown, the set temperature regime does affect the model preferences, and in cases even result in slightly better fits. 

\citet{Patel_2024} suggest that Visit 1 can be explained with an inverted \ce{CO} and \ce{CO2} atmosphere. In our grid, we find equal preference for a non-inverted \ce{N2} atmospheres with $\approx 10^{-3}$ VMR of \ce{HCN}, as well as inverted \ce{PO}-rich models. The latter can have substantial abundances of \ce{CO} and \ce{CO2}; however the requirement for an inverted temperature profile prevents these species from being dominant. Such models require abundant shor-twave absorbers, which can be an array of species that contain P, S, Si and Ti elements. While these models are typically much better fitting than a simple blackbody solution, they have many more parameters and are highly degenerate, making it statistically difficult to identify a single definite solution.

Visit 2 follows a blackbody curve, that can be fit with various models. However, we find no improvement in $\chi^2$ values for any of the compositions.

Visit 3 is less pronounced but, similarly to Visit 1, it resembles an inverted spectrum peaking at 4.5 \textmu m. The scatter in the data results in a large increase of $\chi^2$ values across all models, including blackbody solutions. The shown example is one of the best-fitting cases from our grid, where the emitted feature originates from collision-induced opacity of \ce{N2}-\ce{N2}. However, \ce{N2} is not strictly necessary, as sulphur and carbon-containing atmospheres can also provide suitable fits, given that there is a present temperature inversion. We also see some similarities to Visit 1. Atmospheres with abundant \ce{HCN} may also prove to be a good fit. Consistent with \citet{Patel_2024}, Visit 3 can also be explained with an \ce{SiO}-rich atmosphere. Our models indicate that an inverted \ce{SiO} and \ce{SiS} model is plausible, achieving $ \chi^2 \approx 39$. While Visit 3 also offers no unique solution, it is slightly more tailored towards inverted temperature profiles.

Visit 4 corresponds to the \citet{Hu_2024} NIRCam data, for which we analysed the \texttt{Eureka!} and \texttt{SPARTA} reductions in the previous section. The \texttt{stark} reduction shows a nearly identical fitting preference to \texttt{Eureka!}. The best-fitting model for $f=1/4$ remains the same as shown in \ref{fig:F1}. As with other reductions, \texttt{stark} shows significant degeneracy between models, resulting in similar $ \chi^2$ values for both H-poor and H-rich scenarios.

Visit 5 resembles a flat line that is well fit with \ce{PH3} and \ce{CO} as primary absorbers. Notably, the $\chi^2$ value is improved with increasing \ce{CO} abundance, allowing the spectrum to be fit with just \ce{CO}. Similar behaviour is observed for \ce{PH3}-dominated models. These scenarios assume that the temperature-structure profile is strongly influenced by other atmospheric species that lack significant opacity in the NIRCam wavelength region. As with other visits, multiple scenarios remain plausible explanations, including those rich in \ce{HCN}.

\begin{figure*}
    \centering
        \includegraphics[width=1\textwidth]{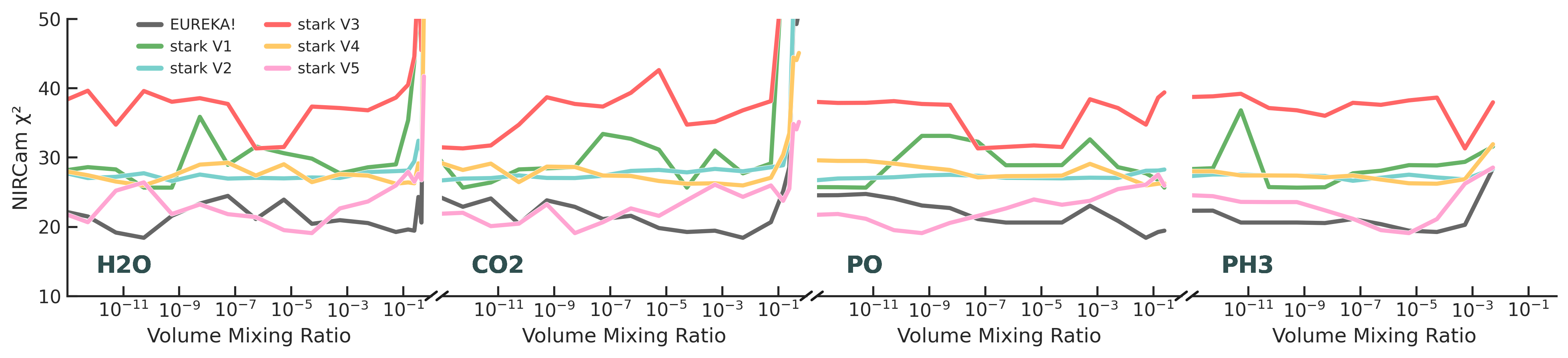}
    \caption{$\chi^2$ dependance on the volume mixing ratio of the denoted species for the \texttt{Eureka!} NIRCam data from \citet{Hu_2024} and for the \texttt{stark} reduction of all five visits from \citet{Patel_2024}. The $\chi^2$ value is the best achieved for each particular abundance summed over the photospheric region ($P>10^{-5}$ bar).}
    \label{fig:F9}
\end{figure*}

\begin{figure*}
    \centering
        \includegraphics[width=1\textwidth]{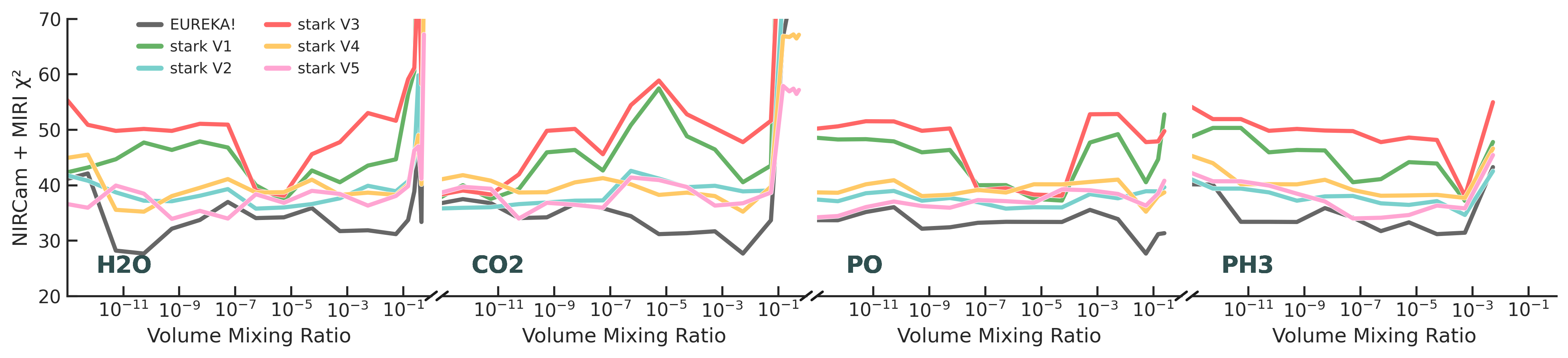}
    \caption{$\chi^2$ dependance on the volume mixing ratio of the denoted species of the combined NIRCam and MIRI spectra for the \texttt{Eureka!} reduction from \citet{Hu_2024} and the \texttt{stark} reduction of the five visits from \citet{Patel_2024}. MIRI data is from \cite{Hu_2024}.The $\chi^2$ value is the best achieved for each particular abundance summed over the photospheric region ($P>10^{-5}$ bar). }
    \label{fig:F10}
\end{figure*}

While model fits of each of the observed 55 Cancri e spectra show markedly different preferences of atmospheric compositions, the overall trends of abundances remain largely unconstrained. In Figure \ref{fig:F9}, we show the $\chi^2$ dependence on the volume mixing ratio of four prominent species: \ce{H2O}, \ce{CO2}, \ce{PO}, and \ce{PH3}. The results are for the \texttt{Eureka!} reduction of NIRCam data from \citet{Hu_2024} and the five \texttt{stark} reductions of NIRCam from \citet{Patel_2024}. A similar analysis for the combined MIRI and NIRCam data is shown in Figure \ref{fig:F10}. Additional trends for the carbon and oxygen mole fractions, as well as the C/O ratio, can be found in Appendix \ref{appendixB}

Similar to \texttt{Eureka!}, the \texttt{stark} spectra show only vague trends in abundances, emphasising the limitations of the current NIRCam data in constraining the atmospheric composition of 55 Cancri e. Visits 3 and 5 have somewhat more distinctive features for the displayed species. For Visit 3, $\chi^2$ is minimised when the \ce{H2O} abundance is between $10^{-7}$ and $10^{-5}$ and the \ce{CO2} abundance becomes vanishingly small. For this visit, a subset of partially inverted, \ce{PH3}-rich models also show notably low $\chi^2$ values.

Visit 5, contains some of the lowest attained $\chi^2$ across the grid, comparable to what we see for the \texttt{Eureka!} reduction. However, its trends generally diverge from \texttt{Eureka!}, except for \ce{PH3}, where both reductions show improved $\chi^2$ towards its increased abundance. For Visit 5, a lower \ce{CO2} abundance also results in improved $\chi^2$, which is a direct consequence of the spectrum aligning more with the \ce{CO} opacity. Despite the differences in the magnitude of $\chi^2$, the trends between the \texttt{Eureka!} and Visit 4 spectra are nearly identical, suggesting that the two pipelines produce consistent spectral features --more so than when compared to the \texttt{SPARTA} reduction

As we have shown previously for the \texttt{Eureka!} reduction, combining the NIRCam data with the MIRI spectrum from \citet{Hu_2024}, similarly improves the trends for the \texttt{stark} reduction. Figure \ref{fig:F10} shows the comparison for the \texttt{stark} spectra. Visits 1 and 3 now exhibit much more distinctive $\chi^2$ variation across all of the indicated species. These visits also become more closely aligned with each other, which is not surprising given their similar inverted spectral shapes. While for Visit 3, the inclusion of MIRI does not alter the preferred atmospheric composition, for Visit 1, the effect is more pronounced, favouring \ce{PH3}-rich models and introducing competing solutions for the \ce{CO2}-rich and \ce{CO2}-poor scenarios. 

As expected, Visit 2, which aligns closely with simple blackbody emission, remains largely unaffected by the addition of MIRI data. 

Interestingly, for \ce{PH3}, all but Visit 5 show improved $\chi^2$ values at higher abundance, with minima occurring near a volume mixing ratio of $10^{-3}$. While the statistical differences are insufficient to constrain the \ce{PH3} abundance robustly, this trend could suggest that the spectral features observed on 55 Cancri e arise from its presence.

Alternatively, the observed spectral differences across all NIRCam visits may point to strong variability in the atmospheric composition, driven by its volatility and large temperature fluctuations. Given the degeneracy between chemical composition and temperature structure, further observations -- preferably at complementary wavelengths —- are required to better constrain the nature of the planet's atmosphere.

\subsection{The effect of surface pressure and heat redistribution}

Although the constraints on the surface pressure of 55 Cancri e's atmosphere are weak, there is some indication that it is surrounded by a substantial global atmosphere \citep{Demory_2016a,Hammond_2017,Angelo_2017,Hu_2024,Patel_2024}. All of our previous results were shown for $P_{surf}=10$ bar, which is consistent with the limits deduced in \citet{Hu_2024,Patel_2024} and is sufficient to support global heat redistribution \citep{Koll_2022}, making this value a logical choice. However, we find that different pressure regimes can significantly affect the model preference and, consequently, the goodness-of-fit of our models. 

\begin{figure}
    \centering
        \includegraphics[width=0.4\textwidth]{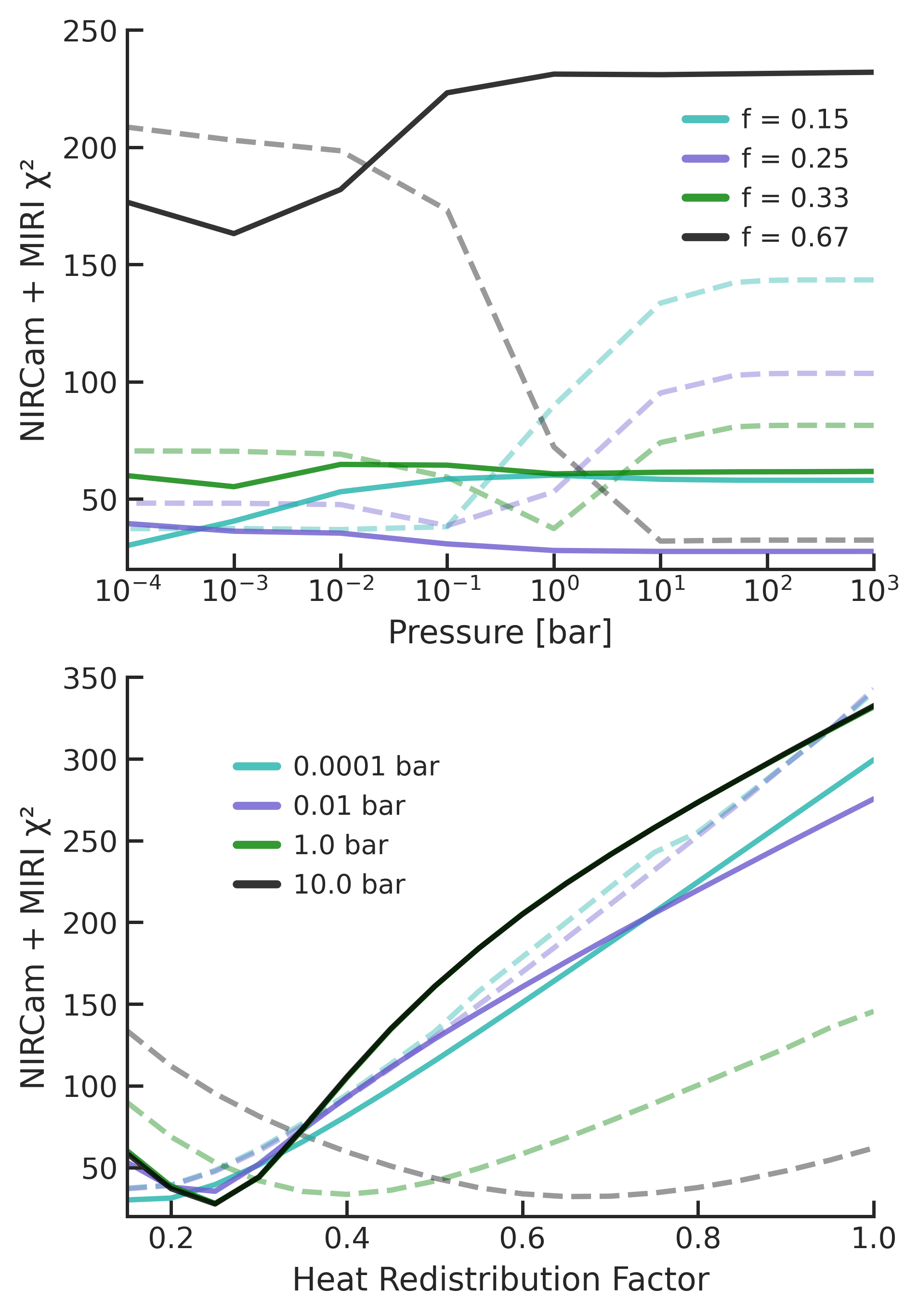}
    \caption{Upper panel: $\chi^2$ dependance on the surface pressure of the atmosphere for two distinct atmospheric compositions. The solid lines represent the \ce{PO}- and \ce{CO2}-rich composition shown in Figure \ref{fig:F1}. Dashed lines are for a \ce{PH3}-rich composition, shown in the upper panel of Figure \ref{fig:F2}. Lower panel: $\chi^2$ dependance on the heat redistribution factor for the same models.}
    \label{fig:F11}
\end{figure}

To demonstrate this, in the top panel of Figure \ref{fig:F11}, we show $\chi^2$ dependance on surface pressure of two distinct compositions. The solid lines represent the best-fitting \ce{PO}- and \ce{CO2}-rich composition from Figure \ref{fig:F1}, with each curve corresponding to a different heat redistribution factor $f$. The $\chi^2$ for this model is minimised when $f=1/4$ and $P \geq 1$ bar. At this temperature regime, decreasing the surface pressure below 1 bar diminishes opacity, which pushes the photospheric region closer to the surface. At $P_{surf}=10^{-2}$ bar, the emitting region aligns closely with the surface, resulting in a worse fit. 

If the heat redistribution factor is $f=0.15$, the changes in atmospheric chemistry and temperature profile result in additional MIR opacity, where the flux decreases below what is required to fit MIRI observations. When major opacities are lost -- which is what happens when the surface pressure falls below 1 bar -- the continuum for $f=0.15$ aligns better with MIRI data, leading to an improvement in $\chi^2$ values. At low temperatures, NIRCam data effectively requires only a small amount of opacity to fit, which is achievable even at very small pressures.

Similar effects, resulting in a bimodal posterior were noted in \citet{Patel_2024}. Depending on the temperature regime, either low-pressure atmospheres (< 1 mbar) or high-pressure atmospheres (> 10 bar) can plausibly explain their NIRCam observations, aligning with what we find for the combined NIRCam and MIRI data.

Increasing the temperature further, to a dayside-confined scenario ($f=2/3$), drastically alters the composition, reducing the dominance of \ce{CO2} and other MIR opacities in favour of short-wavelength counterparts. This results in a strong temperature inversion in the emitting region, misaligning the model continuum with the observations, hence the high overall $\chi^2$ values across all pressures. In this scenario, the $\chi^2$ reaches a minimum only when the surface pressure is low enough for the dominant opacities to vanish ($P_{surf}=10^{-3}$ bar).

An alternative composition with the lowest $\chi^2$ for $f=2/3$ is shown in dashed.  This represents an \ce{N2} atmosphere enriched in H and P, leading to significant abundances of \ce{PH3}, \ce{H2O}, and \ce{NH3}. To fit the MIRI portion of the spectrum at such high temperature (assuming $f=2/3$), a strong continuum is necessary, which arises from abundant \ce{H2O} and \ce{NH3}. At low pressures, the abundances of these species are diminished, whereas at $1 \leq P \leq 10 \, \text{bar}$, the opacities become significant enough to shift the emitting region to the required temperature, minimising $\chi^2$. Increasing the pressure further has no significant effect. This behaviour contrasts with that observed for the primary composition (solid curves). If the alternative composition is shifted to cooler temperature regimes, the continuum no longer aligns with the observed MIRI flux, resulting in a similar trend as found for the primary composition. These two examples demonstrate that the best-fitting surface pressure of the atmosphere is case-dependant and highly degenerate -- not only with the chemical composition but also with the assumed temperature regime.

A similar dependence is observed for the set heat redistribution factor, as shown in the lower panel of Figure \ref{fig:F11}. For the primary model (solid curves), $\chi^2$ is minimised at $f=1/4$, assuming $P_{surf}=10$ bar. The behaviour remains consistent for lower pressures, while pressures above 10 bar have no significant impact on the fit.

For the alternative composition (dashed curves), $\chi^2$ is minimised at $f\approx2/3$. This result stems from the requirement of strong MIR opacity, which the presence of is also less dependant on the atmospheric temperature. As the pressure is decreased, much like for the primary composition, the optimal solution converges towards cooler temperatures, eventually resembling the behaviour of the primary composition. The intuitive conclusion is that at higher baseline temperatures of the emitting region, more substantial opacities are needed to shift the emission to cooler regions that align with the observed flux. When opacities are lost and the spectrum begins to resemble blackbody emission, the $\chi^2$ fit is simply minimised when the continuum aligns with the MIRI data.

\subsection{Variability \& equilibrium with the underlying melt}

Various mechanisms have been proposed to explain the sub-weekly variability observed in the emission of 55 Cancri e. \citet{Loftus_2024} argued that one possible explanation could be a feedback cycle between the atmosphere and the underlying melt. The vaporisation of the melt releases silicate-rich vapour, altering the atmospheric chemistry and potentially triggering cloud formation. This process significantly impacts both the atmospheric temperature structure and the surface temperature of the melt. As a result, the emission spectrum would rapidly vary; if the melt cools, as described in \citet{Loftus_2024}, the influence of silicates diminishes, returning the emission to its initial state. 

\begin{figure*}
    \centering
        \includegraphics[width=0.85\textwidth]{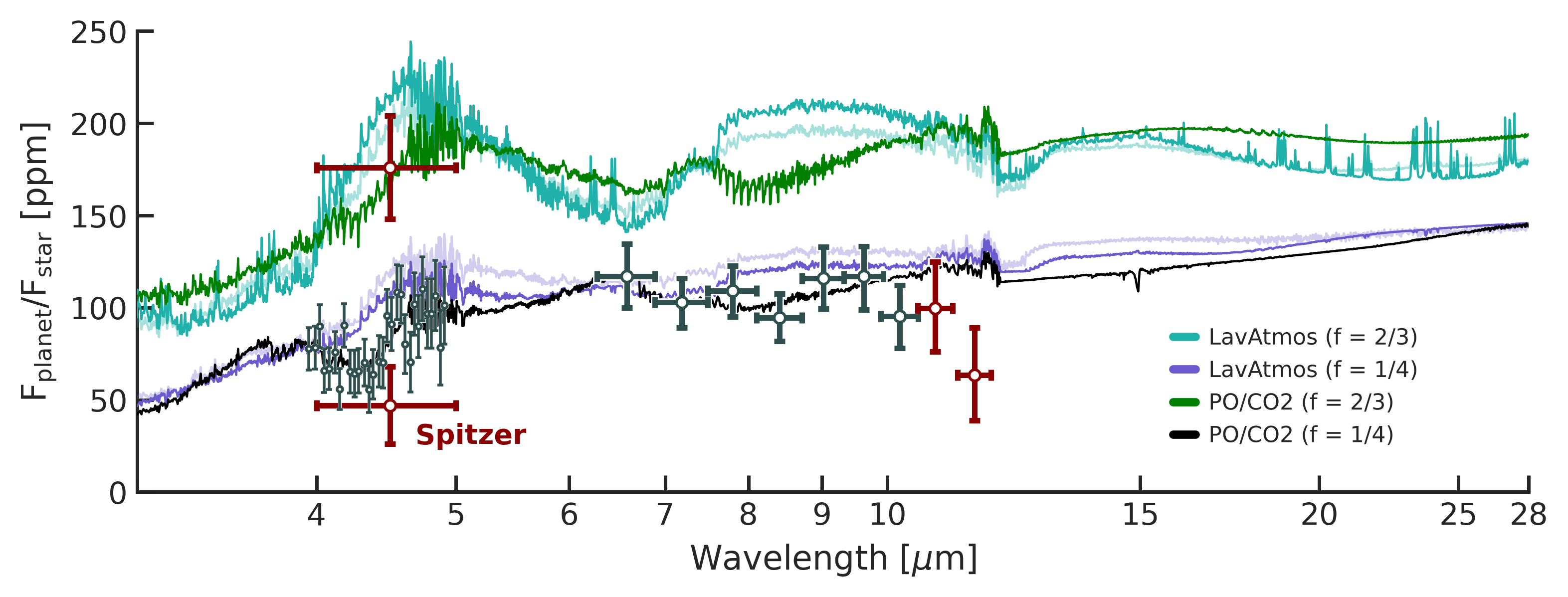}
    \caption{Differences in emission spectra resulting from melt vaporisation into a volatile atmosphere. The black spectrum represents the \ce{PO}- and \ce{CO}-rich atmospheric composition showcased in Figure \ref{fig:F1}, assuming $f=1/4$. The green spectrum corresponds to an identical composition, but for a dayside-confined temperature regime ($f=2/3$). The other coloured spectra illustrate the effects of melt vaporisation for each heat redistribution value, modelled using LavAtmos. Bright spectra represent a 1 bar volatile atmosphere, while faded spectra correspond to a 10 bar volatile atmosphere. The shown JWST data is the \texttt{Eureka!} reduction of the NIRCam and MIRI observations from \citet{Hu_2024}. The two photometric points are the Spitzer data from \citet{Demory_2016a}.}
    \label{fig:F12}
\end{figure*}

The observability of silicate-rich atmospheres has been extensively studied \citep{Schaefer_2009,Miguel_2011,Ito_2015,Zilinskas_2022,Buchem_2022}. However, the interaction with pre-existing volatile envelopes remains an active area of research \citep{Zilinskas_2023,Buchem_2024}. While we do not model condensation and formation of clouds induced by the melt vapour, Figure \ref{fig:F12} shows the influence of melt vaporisation that can be expected to occur 55 Cancri e. The analysis was done using the vaporisation code \texttt{LavAtmos}, which now self-consistently accounts for volatiles in the atmosphere \citep{Buchem_2024}. 

As our base volatile composition, we adopted the \ce{PO}- and \ce{CO2}-rich case (black spectrum), which is also shown in Figure \ref{fig:F1}. Equilibration with the underlying melt introduces a significant influx of silicates, leading to increased abundances of strong absorbers such as \ce{SiS}, \ce{SiO}, and \ce{SiN}. The dominance of \ce{SiO} diminishes the presence of volatile oxides, including key absorbers such as \ce{CO}, \ce{CO2}, and \ce{PO}, drastically altering the features of the emission spectrum. In addition, the introduction of new short-wave opacities creates an inversion in the temperature structure. However, since the degree of silicate vaporisation is strongly dependant on the melt temperature \cite{Buchem_2024}, at $f=1/4$, or equivalently $T_{surf}<2000$ K, the resulting photosphere is nearly isothermal (for purple spectra).

The magnitude of the vaporisation also scales with the initial volatile pressure. The bright purple spectrum corresponds to a 1 bar volatile atmosphere, while the faded purple spectrum represents a 10 bar atmosphere. In this scenario, due to a more inverted temperature structure, a higher initial pressure causes larger differences in the spectral features. For wavelengths probed by NIRCam, the interaction with the melt--even when cloud formation is not included--can already cause substantial variability. The emitted flux depends on the severity of the inversions, which is dictated by the availability of short-wave opacity.

Larger scale variability, as observed by Spitzer \citep{Demory_2016a} and JWST \citep{Patel_2024}, can also be attributed to changes in planet's energy balance. Recent reanalysis of the Spitzer phase curve by \citet{Mercier_2022} suggests an even hotter dayside temperature, which indicates that the atmosphere may be effected by eve stronger temperature inversions. Such changes may arise from variable atmospheric dynamics, atmospheric collapse, cloud formation, or other processes, none of which are explored in this study. If this is the case, the resulting temperature fluctuations would lead to temporary shifts in the emission spectrum, significantly altering the present opacity features.

The green spectrum shows the emission of an identical composition as our baseline model, but under a dayside-confined temperature regime ($f=2/3$). Oscillations between dayside-confined and full heat redistribution regimes are already sufficient to account for the observed extremes. When the volatile envelope is coupled to the hotter melt, we see stronger temperature inversions, resulting in more pronounced spectral features. A high-temperature melt also releases sufficient oxygen to maintain the abundances of \ce{CO}, \ce{CO2}, and \ce{PO}. The introduction of silicates also results in strong \ce{SiO}, \ce{PS}, and \ce{SiS} features. Significant variations in heat redistribution efficiency would likely manifest throughout MIR wavelengths, which are observable by MIRI, emphasising the need for additional JWST observations to investigate these phenomena further.

\subsection{Differences in observability caused by thermodynamics data}
\label{thermo_data}

\begin{figure*}
    \centering
        \includegraphics[width=0.9\textwidth]{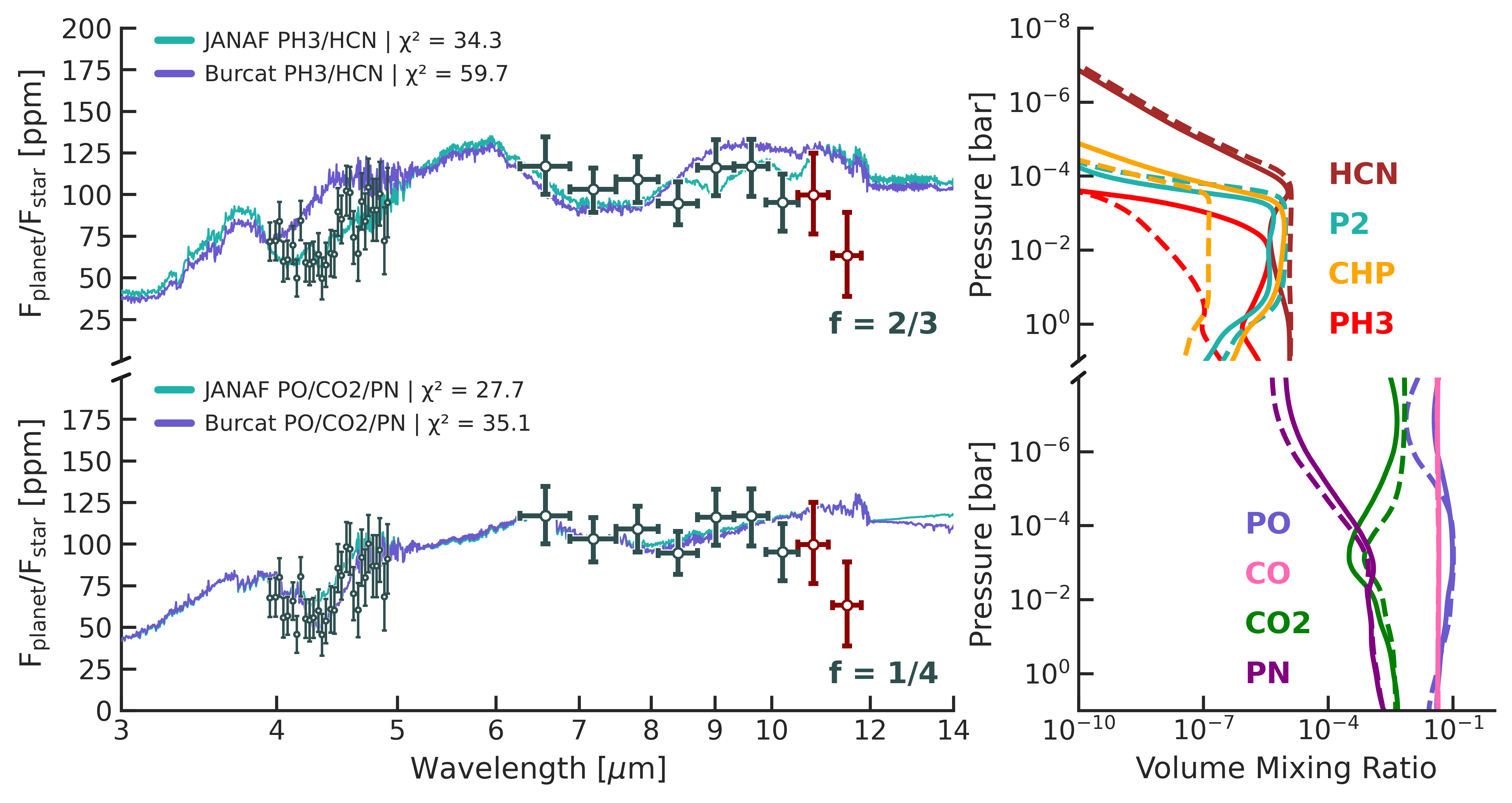}
    \caption{Comparison of emission spectra for models using NIST-JANAF versus Burcat (NASA9) thermodynamic data. The upper panel shows the \ce{PH3}/\ce{HCN} model for $f=2/3$ from Figure \ref{fig:F3}, along with volume mixing ratios of key species. Solid lines represent abundances for the NIST-JANAF models, while dashed lines correspond to the Burcat models. The lower panel shows the \ce{PO}/\ce{CO2} models for $f=1/4$ model from Figure \ref{fig:F1}. Also show is the \texttt{Eureka!} reduction of the NIRCam and MIRI data from \citet{Hu_2024}.}
    \label{fig:F13}
\end{figure*}

To solve equilibrium chemistry, we have utilised the NIST-JANAF thermodynamics database. However, substantial differences exist in the thermodynamic constants and species lists compiled in different databases. In Figure \ref{fig:F13}, we show the spectral differences for two specific cases, highlighting the possible differences in observability when comparing the results from NIST-JANAF and Burcat (NASA9) data. 

In our tested cases, for atmospheres with low phosphorus or silicon content, the differences between these databases are negligible. However, for P- and Si-rich atmospheres, the discrepancies become significant. The upper panel of Figure \ref{fig:F13} shows a \ce{PH3}/\ce{HCN} model, previously presented in Figure \ref{fig:F3}. The major spectrum features in the NIST-JANAF model are from \ce{PH3} and \ce{HCN}. The Burcat model alters the chemistry of P-containing species, effectively eliminating the emission of \ce{PH3}. The reduction in abundances of molecules such as \ce{PH} and \ce{CHP} results in diatomic and monatomic phosphorus the main P-carrier. The resulting spectrum is dominated solely by the \ce{HCN} opacity.

In the lower panel we show the \ce{PO}/\ce{CO2} model, which has the lowest $\chi^2$ value in our grid. In this hydrogen-free atmosphere, \ce{PH3} cannot form. However, differences in the NIRCam portion of the spectrum between NIST-JANAF and Burcat data remain. Variations in the abundances of P-containing species lead to rearrangements in the overall chemistry. For this particular example, the \ce{CO2} volume mixing ratio is increased, making the 4.5 \textmu m feature deeper. This drastically worsens the fit to the observed NIRCam spectrum.

As has been noted in \citet{Lee_2024} and the references therein, thermochemical data is one of the largest uncertainties in forward chemistry models. We compiled our Burcat species list in a similar fashion to \citet{Lee_2024}, with only several new additions. The differences in the chemistry are expected, since Burcat data includes 59 P-containing species, including additional phosphorus hydrides and oxides, compared to just 16 in NIST-JANAF. The extended list features important species such as \ce{HOPO}, \ce{HOPO2}, \ce{HPO}, and \ce{P2O3}, which could potentially be very strong opacity sources. Moreover, as noted by \citet{Lee_2024}, photochemistry may play a critical role in determining the phosphorus chemistry, especially on strongly irradiated lava planets

Although not shown, we observe similar but less pronounced differences for Si-rich chemistry. For P- and S-rich atmospheres, accounting for variations in thermochemical data can shift model preferences. Thus, we caution that the choice of thermodynamic database should be approached carefully, as it may result in significant spectral discrepancies, including the omission of entire features.

\subsection{Improving constraints with future observations}
\label{sec:future_observations}

\begin{figure*}
    \centering
        \includegraphics[width=0.8\textwidth]{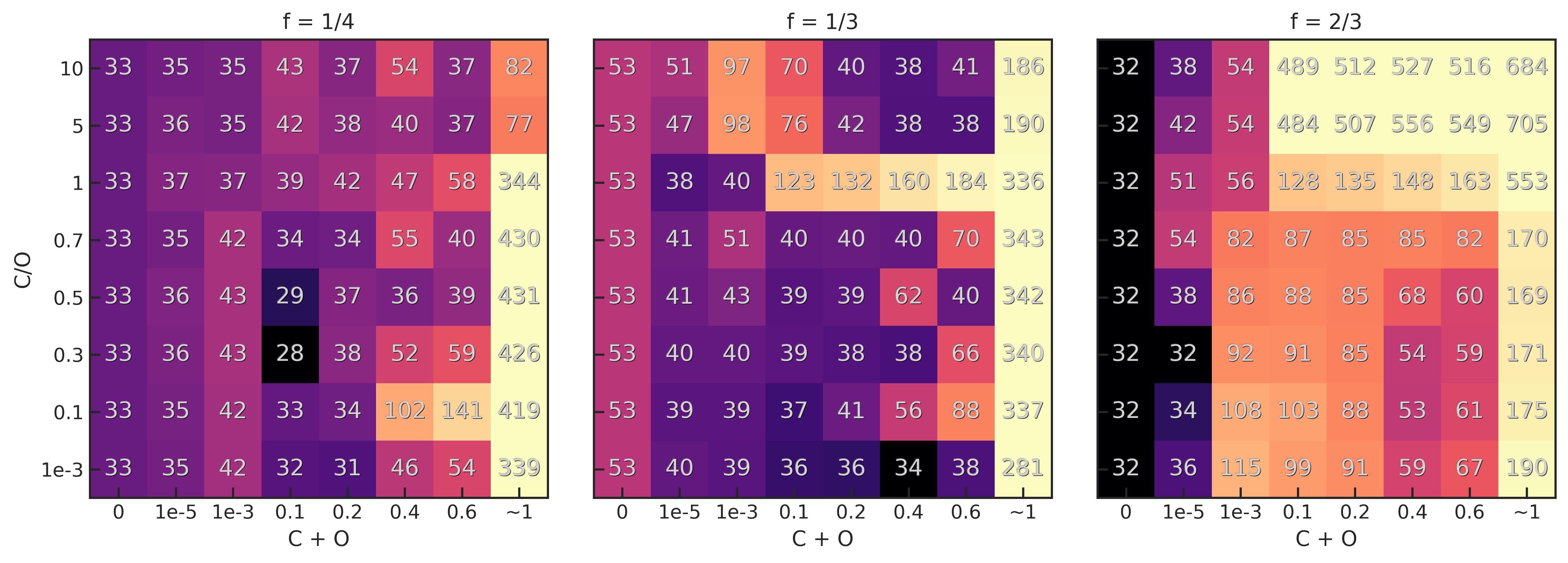}
    \caption{Achieved $\chi^2$ values for varying carbon and oxygen mole fractions in the atmosphere versus the C/O ratio. The results are show for the \texttt{Eureka!} reduction of the combined NIRCam and MIRI spectrum from \citet{Hu_2024}, where the NIRCam flux is not allowed to vary freely and is fixed to a flux offset of +79 ppm. Each panels showcases the results for a different heat redistribution value $f$. The colour gradient and the numbers represent the minimal $\chi^2$.}
    \label{fig:F14}
\end{figure*}

Aside from the unconstrained heat redistribution factor, one of the major sources of degeneracy in the models comes from the undefined absolute flux of the NIRCam data. By fixing the flux offset to our best-fitting model for the \texttt{Eureka!} data from \citet{Hu_2024}, which corresponds to a 79 ppm adjustment, some of the degeneracy between the compositions is mitigated. In Figure \ref{fig:F14}, we show the achieved $\chi^2$ values for varying carbon and oxygen abundances against the C/O ratio, using the \texttt{Eureka!} reduction of the combined NIRCam and MIRI data, now with a fixed NIRCam offset of 79 ppm. 

\begin{figure*}
    \centering
        \includegraphics[width=0.9\textwidth]{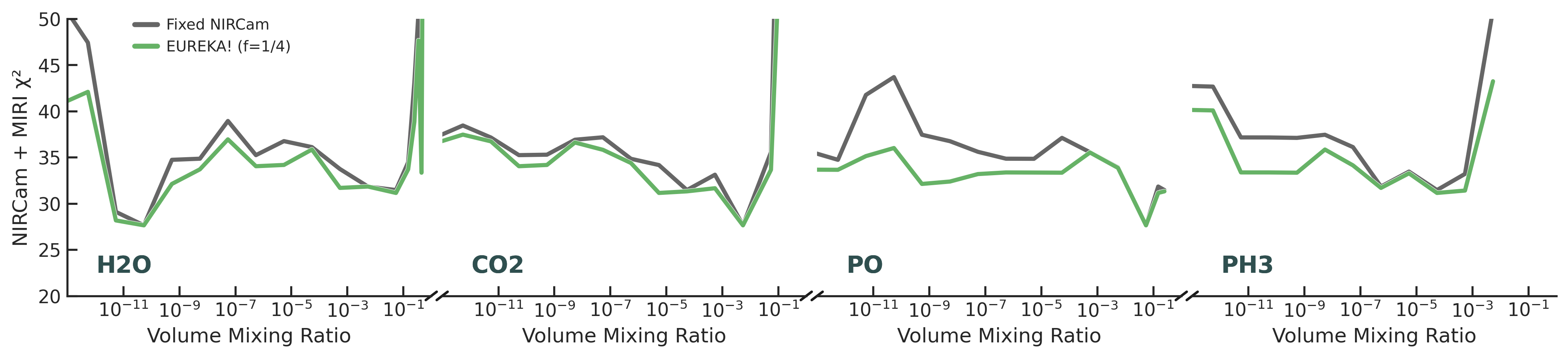}
    \caption{$\chi^2$ dependence on the volume mixing ratios of the indicated species for models with a variable versus fixed NIRCam flux. The models use the combined NIRCam and MIRI \texttt{Eureka!} reduction \citep{Hu_2024} and assume a heat redistribution factor of $f = 1/4$. For each case, species abundances are integrated over the photospheric region ($P > 10^{-5}$ bar)}
    \label{fig:F15}
\end{figure*}

Due to our biased choice towards the cooler temperature regime, for $f=1/4$, we see only a slight improvement in the preference towards the model with C+O = 0.1 and C/O = 0.3. Fixing the absolute flux naturally  increases the $\chi^2$ values for models dominated by strong opacity sources. The effects of this adjustment 
for $f=1/4$ can be more closely seen in the analysis of individual species in Figure \ref{fig:F15}.

At higher temperatures, the convergence towards possible scenarios becomes more distinct. Many of the previously viable compositions are ruled out, leading to much tighter constraints on the atmospheric composition. For $f = 1/3$, atmospheres with either negligible $\text{C+O}$ content or those completely dominated by $\text{C+O}$ exhibit much higher $\chi^2$ values. In the dayside-confined temperature regime ($f = 2/3$), the preference for $\text{C+O}$-free atmospheres becomes even stronger, making reduced, \ce{PH3}-rich compositions a more likely explanation.

\begin{figure*}
    \centering
        \includegraphics[width=0.9\textwidth]{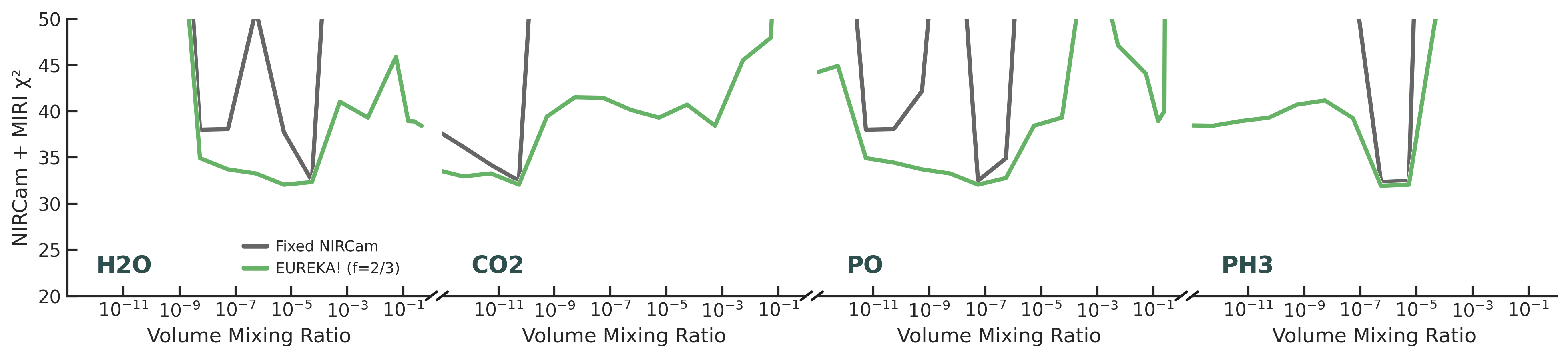}
    \caption{$\chi^2$ dependence on the volume mixing ratios of the indicated species for models with a variable versus fixed NIRCam flux. The models assume a heat redistribution factor of $f = 2/3$. For each case, species abundances are integrated over the photospheric region ($P > 10^{-5}$ bar).}
    \label{fig:F16}
\end{figure*}

In Figures \ref{fig:F15} and \ref{fig:F16}, we compare trends for two scenarios -- fixed versus variable NIRCam flux -- for selected species. Both figures correspond to the combined NIRCam and MIRI datasets from \citet{Hu_2024}. Figure \ref{fig:F15} shows results for $f = 1/4$, while Figure \ref{fig:F16} depicts those for $f = 2/3$. As discussed earlier, the colder regime ($f = 1/4$) shows only slight improvement in abundance trends, with a more notable effect for phosphorus-containing species, further favouring their higher abundances. For the hotter regime ($f = 2/3$, Fig. \ref{fig:F16}), fixing the NIRCam flux results in much stronger constraints, with a clear preference for negligible \ce{CO2} abundance. The \ce{PH3} abundance converges around $10^{-6}$--$10^{-5}$ VMR, consistent with the findings of \citet{Hu_2024}. Similar improvements are observed for \ce{H2O} and \ce{PO}.

It should be noted that the chosen 79 ppm offset introduces a bias towards a specific atmospheric composition and is used here solely to demonstrate the degeneracy caused by an unconstrained NIRCam flux. Determining the absolute flux at 4 -- 5 \textmu m would provide important constraints on the atmospheric composition. However, this remains challenging due to saturation and significant emission variability. Even if the absolute NIRCam flux from the five visits in \citet{Patel_2024} was defined, each visit would favour distinct atmospheric compositions.

\begin{figure*}
    \centering
        \includegraphics[width=0.85\textwidth]{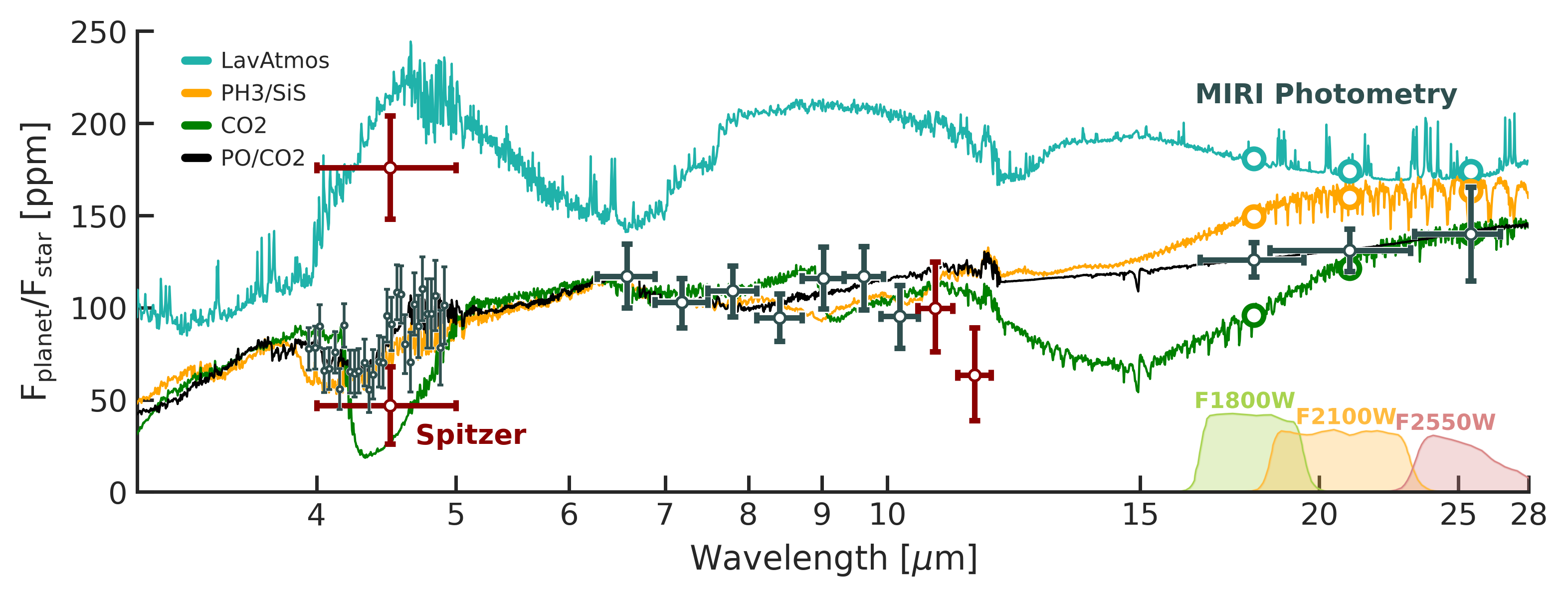}
    \caption{Selection of emission spectra with estimated MIRI photometry noise for 18, 21, and 25 \textmu m. The black spectrum corresponds to the \ce{PO}/\ce{CO2} model from Figure \ref{fig:F1}. The green spectrum is a \ce{CO2}-dominated model. The orange spectrum reflects a \ce{PH3}/\ce{SiS}-rich composition. The turquoise spectrum is a volatile atmosphere that is affected by the vaporisation of the melt, as shown in Figure \ref{fig:F12}, and represent a strong inversion case where the temperature regime is set to $f=2/3$. The Spitzer data is from \citet{Demory_2016a}, and the NIRCam and MIRI data points are for the \texttt{Eureka!} reduction from \citet{Hu_2024}. The calculated MIRI photometric noise corresponds to three eclipses for the F1800W and F2100W filters, and four eclipses for the F2550W filter.}
    \label{fig:F17}
\end{figure*}

Instead, obtaining a thermal phase curve in NIRCam/MIRI or probing longer wavelengths of planetary emission may offer greater insights, particularly in constraining the heat redistribution efficiency and addressing variability, as it remains unknown whether the emission of 55 Cancri e varies at $\geq 5$ \textmu m. In Figure \ref{fig:F17} we show selected emission models alongside the estimated noise for MIRI photometry using the F1800W (18 \textmu m), F2100W (21 \textmu m), and F2250W (25.5 \textmu m) filters. Although MIRI is capable of doing photometry at 15 \textmu m, following the recommended practices listed in the Exposure Time Calculator (ETC) the detector is easily saturated for 55 Cancri e. Among the remaining filters, F1800W achieves the highest precision of 18 ppm. Three eclipses observed at 18 \textmu m could potentially distinguish, at $>3\sigma$, between some of the major atmospheric types, such as \ce{CO2}-dominated models (green) and those with strong temperature inversions. Interaction with the melt may also significantly influence long-wavelength flux, producing substantial, easily detectable deviations from the current JWST data (turquoise model).

\begin{figure}
    \centering
        \includegraphics[width=0.5\textwidth]{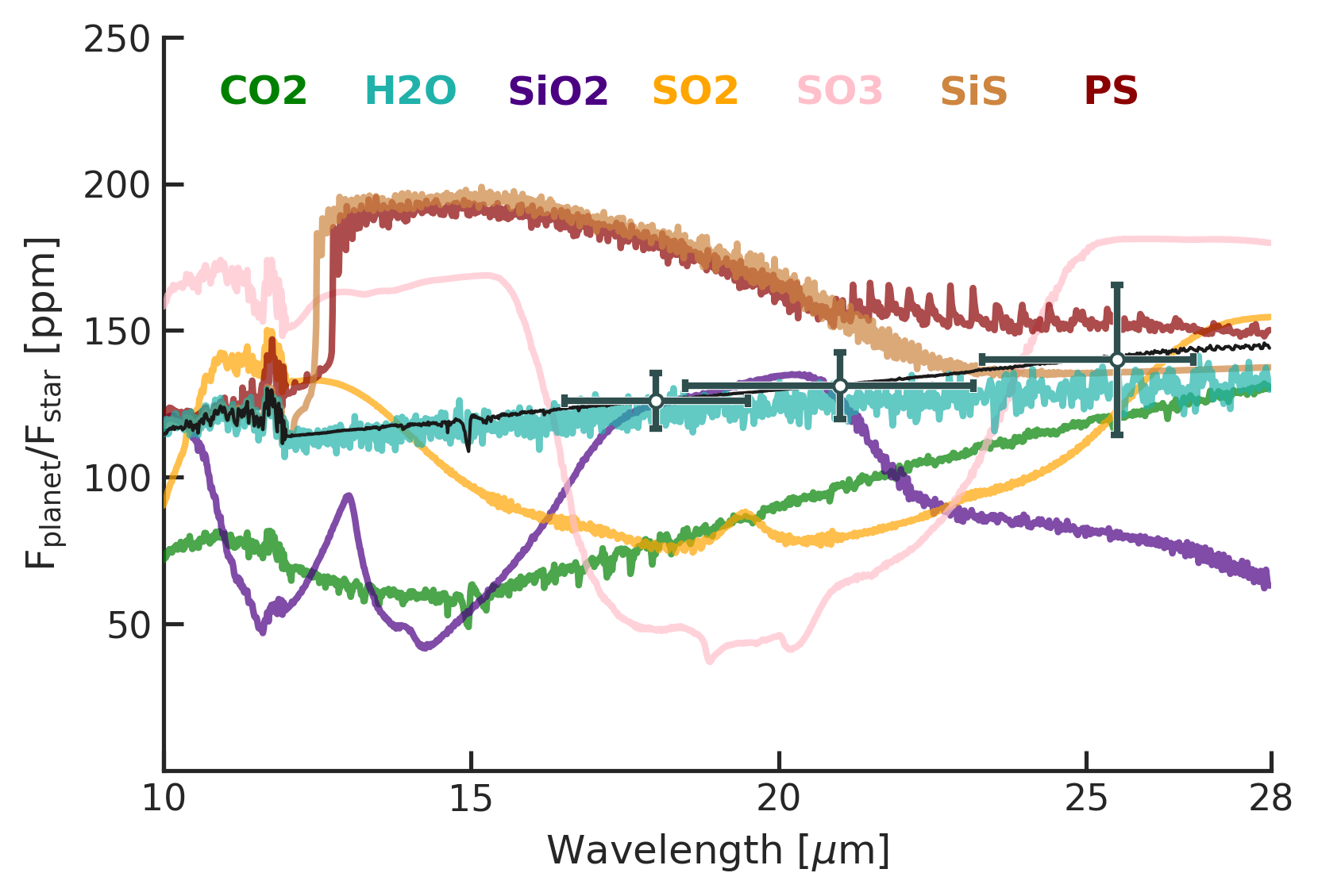}
    \caption{Contribution of individual opacities for the MIRI photometric region. The black spectrum is the \ce{PO}/\ce{CO2} model that is also shown in Figure \ref{fig:F1}. Coloured spectra correspond to models where the indicated species is the dominant opacity contributor. The calculated MIRI photometric noise corresponds to 3 eclipses for the F1800W and F2100W filters, and 4 eclipses for the F2550W filter.}
    \label{fig:F18}
\end{figure}

Using the combined observations from the three photometric points, we can identify trends that deviate from blackbody-like emission. A significant \ce{CO2} opacity would reduce the flux at 18 \textmu m but align with blackbody predictions at longer wavelengths. In inverted temperature scenarios, absorbers like \ce{SiS} or \ce{PS} would create a negative slope. Examples of such individual absorbers are shown in Figure \ref{fig:F18}. For sulphur-dominated atmospheres, we might observe features from \ce{SO2} or \ce{SO3}. Phosphorus-rich atmospheres could exhibit an inverted \ce{PS} feature spanning 13 -- 22 \textmu m. Magma ocean-influenced atmospheres might show the presence of \ce{SiO2} or \ce{SiS}, both of which have substantial opacities at MIR wavelengths. Conversely, \ce{H2O}-rich atmospheres would likely resemble a blackbody continuum and are indistinguishable with MIRI photometric observations. Although mentioned species have prominent opacities, forward models combine multiple contributors, potentially diminishing spectral distinctiveness. Furthermore, many of these features would appear in MIRI/LRS spectra, emphasising the need for repeat JWST observations.

MIRI photometry is a valuable tool for constraining the atmospheric composition of 55 Cancri e and ruling out long-wavelength variability or temperature inversions. However, most best-fitting models lack significant features beyond 18 \textmu m, making it unclear whether the observed flux originates from a continuum opacity or a low-temperature blackbody emission. When combined with the \texttt{Eureka!} reduction of NIRCam and MIRI, some improvements in atmospheric abundance constraints are observed, but these are less significant than fixing the 4–5 \textmu m flux. This underscores the necessity of not only the need of acquiring more data but also obtaining higher-quality data with continuous wavelength coverage and improved precision.

\section{Discussion}
\label{sec:Discussion}

Characterising extreme planets, such as 55 Cancri e, presents significant challenges. Much of our knowledge of it is extrapolated from better-understood exoplanets, which may not capture the diversity in atmospheric chemistry and dynamics possible on highly irradiated rocky planets. The atmospheres surrounding these worlds may reveal greater diversity in chemistry than is currently expected. 

In our study, we employed a forward model grid search to analyse the observed spectrum of 55 Cancri e. This approach contrasts with retrieval analyses, which are frequently used to interpret exoplanet observations. Retrievals, while powerful, rely on simplifying assumptions. They decouple the temperature structure from the chemistry by applying basic parametrisation that prioritises statistical fitting over physical consistency. While this framework is effective for well-understood systems, assumptions introduced by our choices of priors and limited compositions can result in misleading conclusions. 

For example, the NIRCam visit presented in \citet{Hu_2024} and the Visit 4 in \citet{Patel_2024} are the same data processed using different data reduction pipelines. The final spectra are only slightly different. However, the derived conclusions based on spectral retrievals between these data analyses are vastly different, which may be due to subtle differences in the data reduction and the assumptions made in the spectral retrievals. Here, using grids of self-consistent models, we show that the subtle differences in the data analysis does not result in substantial difference in the preference for a CO2- and PO-rich atmosphere.

A grid search of self-consistent forward models enforces consistency between chemical equilibrium and temperature structure while also allowing for the exploration of complex and uncharted scenarios. This is particularly important for close-in rocky planets, where atmospheric regimes may be far removed from typical planetary environments. High surface temperatures can partially vaporise their crusts resulting in abundant melt refractories being released into the atmosphere \citep{Schaefer_2009,Miguel_2011,Ito_2015,Zilinskas_2022,Buchem_2022}. Increasing temperatures can also allow for sulphur and phosphorus components to potentially become important constituents \citep{Schaefer_2012,Herbort_2020}. Depending on the melt composition, interior and atmospheric dynamics, surface activity, and the degree of chemical disequilibrium, geochemically predicted \ce{CO2}/\ce{CO} or \ce{N2} atmospheres may be substantially altered by the exotic chemistry involving, for example, various phosphorus, sulphur, or silicon species \citep{Omran_2021,Boukar_2023,Meier_2023b,Byrne_2024,Buchem_2024,Lee_2024,Tian_2024}.

By systematically exploring a large, physically motivated parameter space of varying compositions, surface pressures, and heat redistribution efficiencies, a grid search can comprehensively evaluate the interplay between different atmospheric scenarios, providing an overview and insights that retrievals might simply overlook. However, forward modelling is computationally intensive, as it requires generating and evaluating thousands of self-consistent models. Retrieval analyses, by comparison, are computationally efficient and excel at quantifying uncertainties and degeneracies within the defined parameter space. As our understanding of atmospheric chemistry and opacity sources expands, the computational gap between the two methods used to analyse exotic atmospheres may narrow. Currently, the choice of method depends on the complexity of the system being studied and the breadth of the parameter space required to explain the observations. Ideally, forward models and retrievals should be seen as complementary tools, with forward models defining physically motivated priors for retrievals and retrievals refining insights provided by forward models.

Our results demonstrate that, even with extensive forward modelling, there are significant degeneracies between the chemistry and the properties of the atmosphere. For instance, atmospheres dominated by \ce{CO2} or \ce{H2O} can mimic spectral signatures of \ce{PH3}-rich scenarios, largely due to the temperature structure defining emission features. This large-scale of degeneracy simply indicates that the current observational data is not sufficient to characterise 55 Cancri e.

Continuous coverage and high-precision future observations could provide the necessary constraints to characterise 55 Cancri e. However, the development of efficient, adaptive grid-search techniques that balance computational feasibility with physical rigour will be essential for advancing our understanding of such extreme exoplanets. For now, the combination of forward modelling and retrieval analyses offers the best pathway to uncovering the diversity of exoplanets.

\section{Conclusion}
\label{sec:Conclusions}

JWST observations of 55 Cancri e have provided new insights into the planet's potential atmosphere, yet large degeneracies between models prevent us from robustly constraining its composition or dominant absorbers. Current observations suggest the presence of a thick volatile envelope, potentially rich in \ce{CO2} or \ce{CO}. This inference comes from the low observed flux in the MIRI/LRS spectrum and a tentative absorption feature in the NIRCam spectrum \citep{Hu_2024}. However, the forward model analysis by \citet{Hu_2024} revealed a high level of degeneracy, demonstrating that the data can also be explained by a variety of alternative compositions. Meanwhile, additional NIRCam observations by \citet{Patel_2024} show strong sub-weekly variability, introducing further uncertainty.

Building on the results of \citet{Hu_2024}, we constructed a grid of over 25000 self-consistent forward models, incorporating H-N-O-C-S-P-Si-Ti equilibrium chemistry, to assess plausible atmospheric compositions. We conducted a goodness-of-fit analysis for the combined and individual NIRCam and MIRI spectra, accounting for various data reduction approaches from \citet{Hu_2024,Patel_2024}. For the spectra presented by \citet{Hu_2024}, our results suggest that the planet is consistent  with a global $\geq10$ bar nitrogen atmosphere dominated by \ce{CO2} and \ce{PO} that is free of hydrogen. However, many alternative compositions, such as \ce{H2O}-, \ce{PH3}-, and \ce{Si}-rich scenarios, show statistically comparable fits. We also find that the heat redistribution factor and surface pressure are highly degenerate with atmospheric composition, complicating efforts to constrain these parameters independently. Additionally, we have shown that the predicted emission spectra for P- and Si-rich compositions are strongly influenced by the chosen thermodynamic data, in cases resulting in completely erased features.

The highly variable NIRCam observations by \citet{Patel_2024} reveal distinct spectral preferences across visits, suggesting diverse atmospheric scenarios, including \ce{HCN}- and \ce{PH3}-rich compositions, as well as inverted temperature profiles driven by opacities from \ce{CO2}, \ce{CO}, or even \ce{N2-N2} collision-induced absorption.  Combining these NIRCam data with the MIRI spectrum from \citet{Hu_2024} reduces some of the degeneracies but does not significantly constrain overall composition trends. As we have shown, the observed variability may stem from interactions between the atmosphere and an underlying magma ocean, driving rapid changes in atmospheric chemistry and emission flux.

One of the largest sources of uncertainty in the analysis is the unconstrained absolute flux in the NIRCam observations and the unconstrained heat redistribution factor. Our models suggest that determining these values through follow-up observations could significantly improve compositional constraints.

While rapid variability in the planet's emission complicates the efforts to retrieve 4 -- 5 \textmu m emission flux, phase curve observations and JWST MIRI photometry at longer wavelengths (18–25 \textmu m), could prove more useful. Long wavelength photometry would also help address variability and confirm the presence or absence of a temperature inversion. Although while these wavelengths contain many identifiable opacities, many of our best-fitting models with combined opacities predict diminished spectral features. Such models still have substantial features in the MIRI/LRS region, emphasising the need for high-precision and continuous spectral coverage at $\geq 5$ \textmu m. 

In conclusion, 55 Cancri e remains one of the most challenging yet critical targets for atmospheric characterisation and planetary evolution studies. While current data provide compelling evidence for a volatile atmosphere on a rocky exoplanet, they also highlight the need for further observations that would refine our understanding of this unique planet.

\begin{acknowledgements}
 
\\
M.Z., Y.M and L.J.J. acknowledge funding from the European Research Council (ERC) under the European Union’s Horizon 2020 research and innovation programme (grant agreement no. 101088557, N-GINE). C.P.A.B. acknowledges support through a Dutch Science Foundation (NWO) Planetary and Exoplanetary Science (PEPSci) grant. S.Z. was supported by NASA through the NASA Hubble Fellowship grant \#HST-HF2-51570.001-A awarded by the Space Telescope Science Institute, which is operated by the Association of Universities for Research in Astronomy, Incorporated, under NASA contract NAS5- 26555. E.S. acknowledges support from the Dutch Research Council (NWO) under the grant OCENW.M.21.264. Part of this research was carried out at the Jet Propulsion Laboratory, California Institute of Technology, under a contract with the National Aeronautics and Space Administration (80NM0018D0004). This work was done as an outside activity and not in the author's capacity as an employee of the Jet Propulsion Laboratory, California Institute of Technology.
\\
\\
Software used in this work:
\texttt{HELIOS-K} \citep{Grimm_2015,Grimm_2021}; \texttt{HELIOS}\citep{Malik_2017,Malik_2019}; \texttt{FASTCHEM} \citep{Stock_2018}; \texttt{LavAtmos} \citep{Buchem_2022}; \texttt{petitRADTRANS} \citep{Molliere_2019,Molliere_2020}.
\\
\\
Any additional supplementary material is available on request from the author.

\end{acknowledgements}
\bibliographystyle{aa}
\bibliography{references}
\onecolumn
\begin{appendix} 

\section{Opacity data}
\label{appendixA}
In Table \ref{table:opacities} we show a compilation of sources for the opacities and line lists used in this study. For \texttt{HELIOS}, we used the precomputed opacity database DACE\footnote{https://dace.unige.ch/} and, where necessary, the opacity calculator \texttt{HELIOS-K}\footnote{https://github.com/exoclime/HELIOS-K} \citep{Grimm_2015,Grimm_2021} to compute the opacities from line lists. To calculate the final spectrum with \texttt{petitRADTRANS} we used the precomputed opacities from the ExoMol\footnote{https://www.exomol.com/} database \citep{Chubb_2021}, using the same line lists where possible. For cases where the precomputed opacity was unavailable, we converted \texttt{HELIOS} opacities to the petitRADTRANS format. Line lists from the Kurucz \citep{Kurucz_1992}, VALD \citep{Ryab_2015}, and HITRAN \citep{Gordon_2017} databases were extensively used. Specific details are provided in the accompanying table.

\begin{longtable}{lllll} 
 \caption[]{Description of the opacities used to calculate temperature profiles and emission spectra.}
 \\
 \label{table:opacities}
Species & Source & Line list & Line List Reference
    \\
  \hline
  \hline
  \\
  \ce{Al} & DACE & VALD & \citet{Ryab_2015}\\
  \ce{AlH} & HELIOS-K & AlHambra & \citet{Yurchenko_2018b}\\
  \ce{AlO} & HELIOS-K & ATP & \citet{Patrascu_2015} \\
  \ce{C} & DACE & Kurucz & \citet{Kurucz_1992} \\
  \ce{C2} & DACE & 8states & \citet{Yurchenko_2018d} \\
  \ce{C2H2} & DACE & aCeTY & \citet{Chubb_2020}\\
  \ce{C2H4} & DACE & MaYTY & \citet{Mant_2018}\\  
  \ce{Ca} & DACE & VALD & \citet{Ryab_2015}\\
  \ce{CaH} & HELIOS-K & MoLLIST & \citet{Li_2012,Bernath_2020}\\
  \ce{CaO} & HELIOS-K & VBATHY & \citet{Yurchenko_2016}\\
  \ce{CaOH} & DACE & OYT6 & \citet{Owens_2022}\\
  \ce{CH} & DACE & MoLLIST & \citet{Masseron_2014,Bernath_2020}\\
  \ce{CH3} & DACE & AYYJ & \citet{Adam_2019}\\
  \ce{CH4} & DACE & YT34to10 & \citet{Yurchenko_2017}\\
  \ce{CN} & HELIOS-K & Trihybrid & \citet{Syme_2021}\\  
  \ce{CO} & DACE & Li2015 & \citet{Gang_2015}\\
  \ce{CO2} & DACE & HITEMP \& UCL-4000$^c$ & \citet{Rothman_2010,Yurchenko_2020}\\
  \ce{CS} & DACE & JnK & \citet{Paulose_2015}\\ 
  \ce{Fe} & DACE & VALD & \citet{Ryab_2015}\\
  \ce{FeH} & DACE & MoLLIST & \citet{Dulick_2003}\\
  \ce{H2+} & DACE & ADJSAAM & \citet{Amaral_2019}\\
  \ce{H2CO} & DACE & AYTY & \citet{Refaie_2015}\\
  \ce{H2O} & DACE & POKAZATEL & \citet{Polyansky_2018}\\  
  \ce{H2O2} & DACE & APTY & \citet{Refaie_2016}\\  
  \ce{H2S} & DACE & AYT2 & \citet{Azzam_2016}\\  
  \ce{H3O+} & DACE & eXeL & \citet{Yurchenko_2020}\\  
  \ce{HCN} & HELIOS-K & Harris & \citet{Barber_2014}\\  
  \ce{HNO3} & DACE & AIJS & \citet{Pavlyuchko_2015}\\  
  \ce{HS} & HELIOS-K & GYT & \citet{Gorman_2019}\\  
  \ce{K} & DACE & VALD & \citet{Ryab_2015}\\
  \ce{KOH} & DACE & OYT4 & \citet{Owens_2021}\\
  \ce{Mg} & DACE & Kurucz & \citet{Kurucz_1992}\\
  \ce{MgH} & HELIOS-K & MoLLIST & \citet{GharibNezhad_2013,Bernath_2020}\\    
  \ce{MgO} & HELIOS-K & LiTY & \citet{Li_2019}\\
  \ce{N} & DACE & VALD & \citet{Ryab_2015}\\
  \ce{N2} & DACE & WCCRMT & \citet{Western_2018}\\
  \ce{N2O} & DACE & HITEMP2019 & \citet{Hargreaves_2019}\\
  \ce{Na} & DACE & VALD & \citet{Ryab_2015}\\
  \ce{NaH} & HELIOS-K & Rivlin & \citet{Rivlin_2015}\\
  \ce{NaO} & DACE & NaOUCMe & \citet{Mitev_2022}\\
  \ce{NaOH} & DACE & OYT5 & \citet{Owens_2021}\\
  \ce{NH} & DACE & MoLLIST & \citet{Fernando_2018}\\
  \ce{NH3} & DACE & CoYuTe & \citet{Coles_2019}\\ 
  \ce{NO} & DACE & XABC & \citet{Wong_2017,Qu_2021}\\ 
  \ce{NO2} & DACE & HITEMP2019 & \citet{Hargreaves_2019}\\ 
  \ce{NS} & DACE & SNaSH & \citet{Yurchenko_2018c}\\ 
  \ce{OH} & DACE & HITEMP & \citet{Rothman_2010}\\  
  \ce{OH+} & DACE & MoLLIST & \citet{Hodges_2017}\\    
  \ce{P} & DACE & VALD & \citet{Ryab_2015}\\    
  \ce{P2H2} & DACE & OY-trans & \citet{Owens_2019}\\    
  \ce{PC} & DACE & MoLLIST & \citet{Ram_2014,Qin_2021}\\    
  \ce{PH} & HELIOS-K & LaTY & \citet{Langleben_2019}\\ 
  \ce{PH3} & DACE & SAlTY & \citet{Sousa_2015}\\ 
  \ce{PN} & DACE & YYLT & \citet{Yorke_2014}\\ 
  \ce{PO} & DACE & POPS & \citet{Prajapat_2017}\\ 
  \ce{PS} & HELIOS-K & POPS & \citet{Prajapat_2017}\\ 
  \ce{S} & DACE & VALD & \citet{Ryab_2015}\\ 
  \ce{SH} & DACE & GYT & \citet{Gorman_2019}\\ 
  \ce{Si} & DACE & VALD & \citet{Ryab_2015}\\
  \ce{SiH} & HELIOS-K & Sightly & \citet{Yurchenko_2018a}\\  
  \ce{SiH2} & HELIOS-K & CATS & \citet{Clark_2020}\\  
  \ce{SiH4} & DACE & OY2T & \citet{Owens_2017}\\  
  \ce{SiN} & DACE & SiNfull & \citet{Semenov_2022}\\  
  \ce{SiO} & HELIOS-K & SiOUVenIR & \citet{Yurchenko_2022}\\
  \ce{SiO2} & DACE & OYT3 & \citet{Owens_2020}\\
  \ce{SiS} & DACE & UCTY & \citet{Upadhyay_2018}\\
  \ce{SO} & DACE & SOLIS & \citet{Brady_2024}\\
  \ce{SO2} & ExoMol & ExoAmes & \citet{Underwood_2016a}\\ 
  \ce{SO3} & ExoMol & UYT2 & \citet{Underwood_2016b}\\     
  \ce{Ti} & DACE & VALD & \citet{Ryab_2015}\\
  \ce{TiH} & HELIOS-K & MoLLIST & \citet{Burrows_2005,Bernath_2020}\\
  \ce{TiO} & HELIOS-K & Toto & \citet{McKemmish_2019}\\
\\
    Scattering and Continuum \\
  \hline    
  \\
  \ce{CO}, \ce{CO2}, \ce{CH4}, \ce{H} &  & Scattering \\    
  \ce{H2}, \ce{H2O}, \ce{N2}, \ce{O2} &  & Scattering \\
  \ce{H-} &  & Continuum (bf \& ff)  & \citet{John_1988,Gray_2008}\\  
  \ce{H2-H2} & \texttt{petitRADTRAN} & CIA & \citet{Borysow_2001,Borysow_2002}\\ 
  \ce{CO2-CO2} & \texttt{petitRADTRAN} & CIA & \citet{Karman_2019}\\ 
  \ce{O2-O2} & \texttt{petitRADTRAN} & CIA & \citet{Karman_2019}\\ 
  \ce{N2-H2} & \texttt{petitRADTRAN} & CIA & \citet{Karman_2019}\\   
  \ce{N2-N2} & \texttt{petitRADTRAN} & CIA & \citet{Karman_2019}\\   
  \\
  \hline
  \multicolumn{4}{p{17.4cm}}{\footnotesize$^{c}$ We use HITEMP2010 for temperature profiles and UCL-4000 for emission spectra; \footnotesize$^{d}$.}
\\

\end{longtable}

\section{Additional analysis of statistical preference for elemental abundances and ratios for the \texttt{Eureka!}, \texttt{SPARTA}, and \texttt{stark} reductions}
\label{appendixB}

\subsection{Individual NIRCam and MIRI C+O and C/O trends for the \texttt{Eureka!} reduction}

\begin{figure*}[h!]
    \centering
        \includegraphics[width=0.8\textwidth]{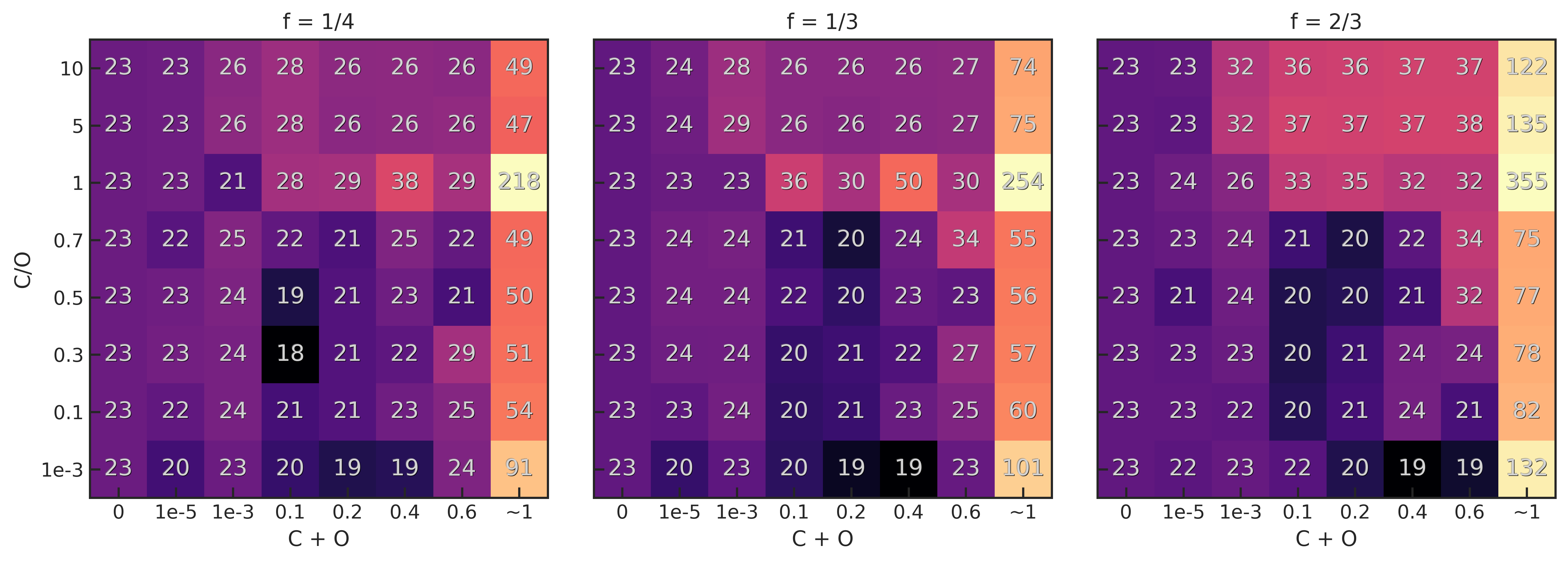}
    \caption{Minimum $\chi^2$ values for varying carbon and oxygen mole fractions as a function of the C/O ratio for the \texttt{Eureka!} reduction of the NIRCam spectrum from \citet{Hu_2024}. Each panel corresponds to a specific temperature regime, indicated by the $f$ value above. The colour gradient and numerical labels represent the minimum $\chi^2$ values for each case.}
    \label{fig:A1}
\end{figure*}

Figure \ref{fig:A1} displays the $\chi^2$ dependence on the mole fractions of C+O and the C/O ratio for the \texttt{Eureka!} reduction of the NIRCam spectrum from \citet{Hu_2024}. Unlike Figure \ref{fig:F6} in the main text, this analysis includes only the NIRCam data, excluding the MIRI spectrum.

The modulation of the NIRCam spectrum is consistent with \ce{CO2} opacity,  which is why the best fitting cases remain to be favourable for its formation. However, in contrast to the combined spectrum, we do see improved $\chi^2$ values for higher C+O mole fractions. This trend is observed across all heat redistribution regimes.  The unconstrained absolute flux of NIRCam is what allows the data to accommodate various atmospheric compositions irrespective of the temperature regime.

\begin{figure*}[h!]
    \centering
        \includegraphics[width=0.8\textwidth]{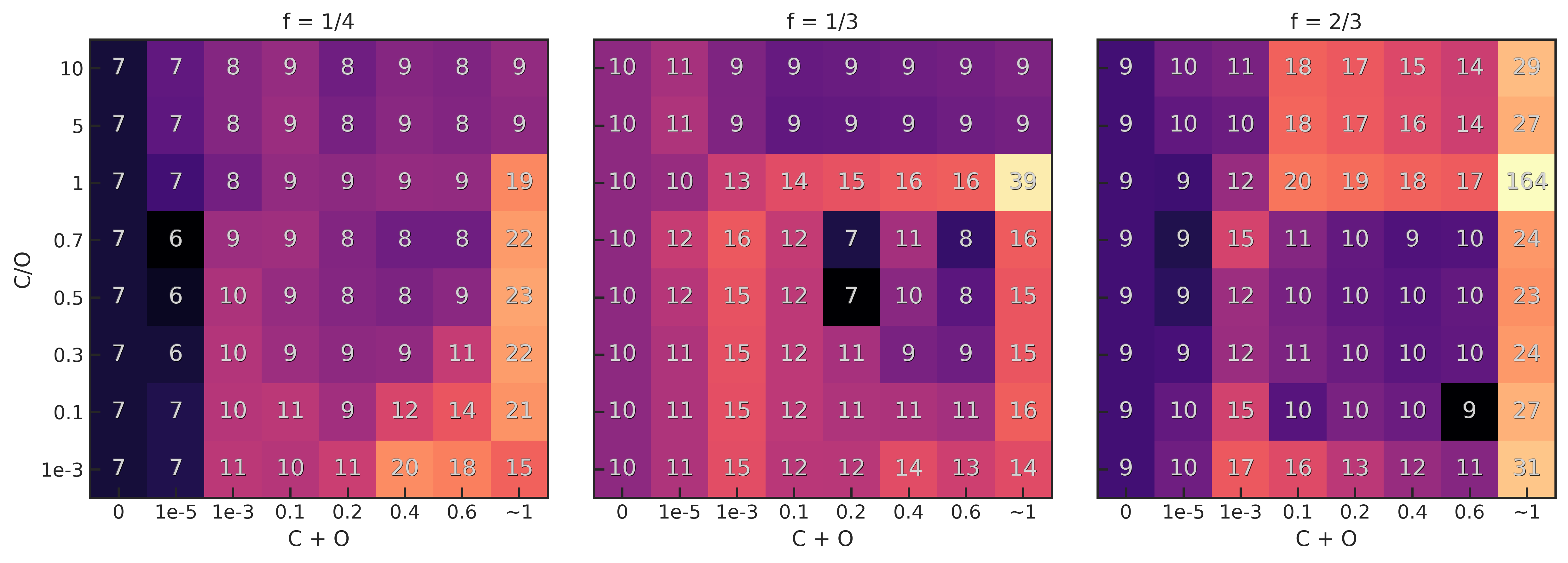}
    \caption{$\chi^2$ values for varying carbon and oxygen mole fractions versus the C/O ratio for \texttt{Eureka!}'s reduction of the MIRI spectrum from \citet{Hu_2024}. Each panel corresponds to a specific temperature regime, indicated by the $f$ value above. The colour gradient and numerical labels represent the minimum $\chi^2$ values for each case.}
    \label{fig:A2}
\end{figure*}

In Figure \ref{fig:A2}, we present the results for the low-resolution MIRI spectrum, which is predominantly featureless and consistent with a blackbody model. At cooler temperatures ($f=1/4$), the spectrum shows a slight preference for compositions devoid of carbon and oxygen, as high abundances of \ce{CO2} or \ce{CO} can lower the flux below observed levels. As the temperature increases, the flux continuum exceeds the observations, requiring strong IR opacity to achieve a good fit. Consequently, $\chi^2$ values are minimised at higher C+O mole fractions, where opacities from species such as \ce{CO2} and \ce{H2O} contribute to improved fits. At dayside-confined temperatures, even greater opacity is necessary to reconcile the low observed flux, leading to significant degeneracies. Here, a wide range of compositions, including carbon-, hydrogen-, and phosphorus-rich chemistries, are viable. Notably, the MIRI spectrum can be adequately described by a low-flux blackbody continuum, which achieves a minimum $\chi^2$ value of 9.4, in contrast to 6.3 attained by our grid of models.

\subsection{C+O and C/O trends for the \texttt{SPARTA} reduction}

\begin{figure*}[h!]
    \centering
        \includegraphics[width=0.8\textwidth]{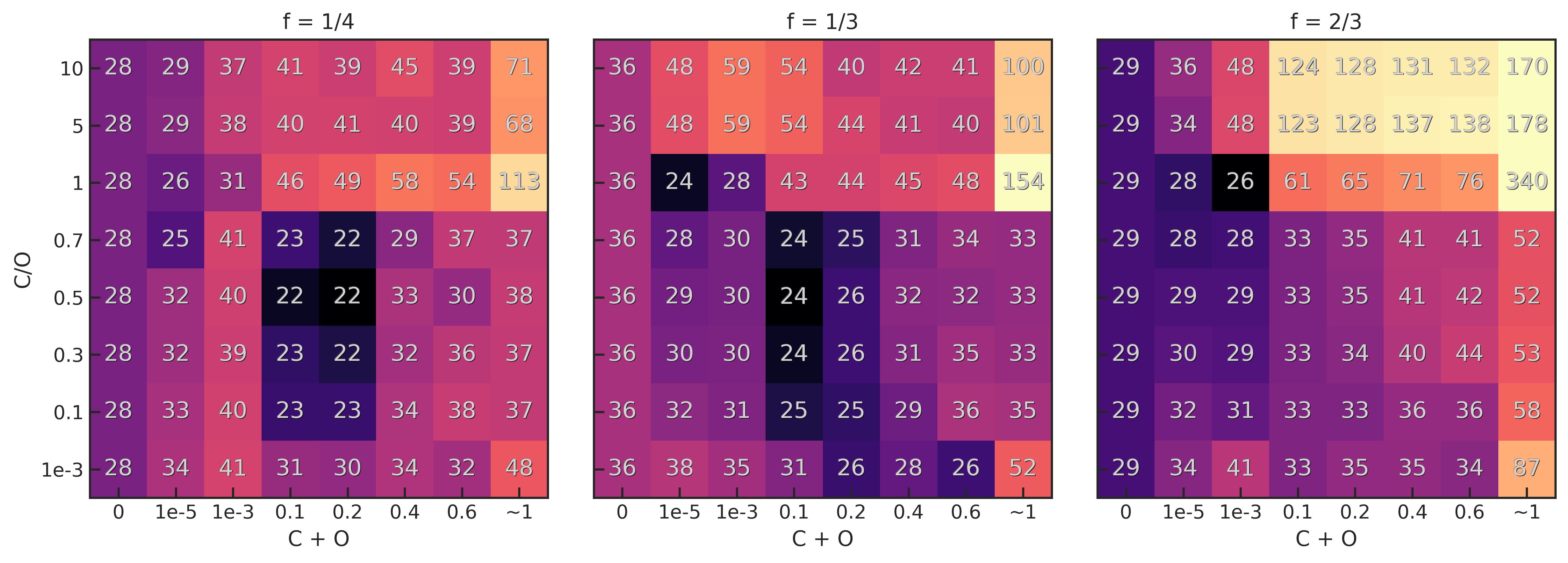}
    \caption{Analysis of the full NIRCam + MIRI \texttt{SPARTA} reduction from \citet{Hu_2024}, showing $\chi^2$ trends for varying carbon and oxygen mole fractions against the C/O ratio. Each panel corresponds to a specific temperature regime, indicated by the $f$ value above it. The colour gradient and numbers denote the minimum $\chi^2$ values for each case.}
    \label{fig:A3}
\end{figure*}

\begin{figure*}[h!]
    \centering
        \includegraphics[width=0.8\textwidth]{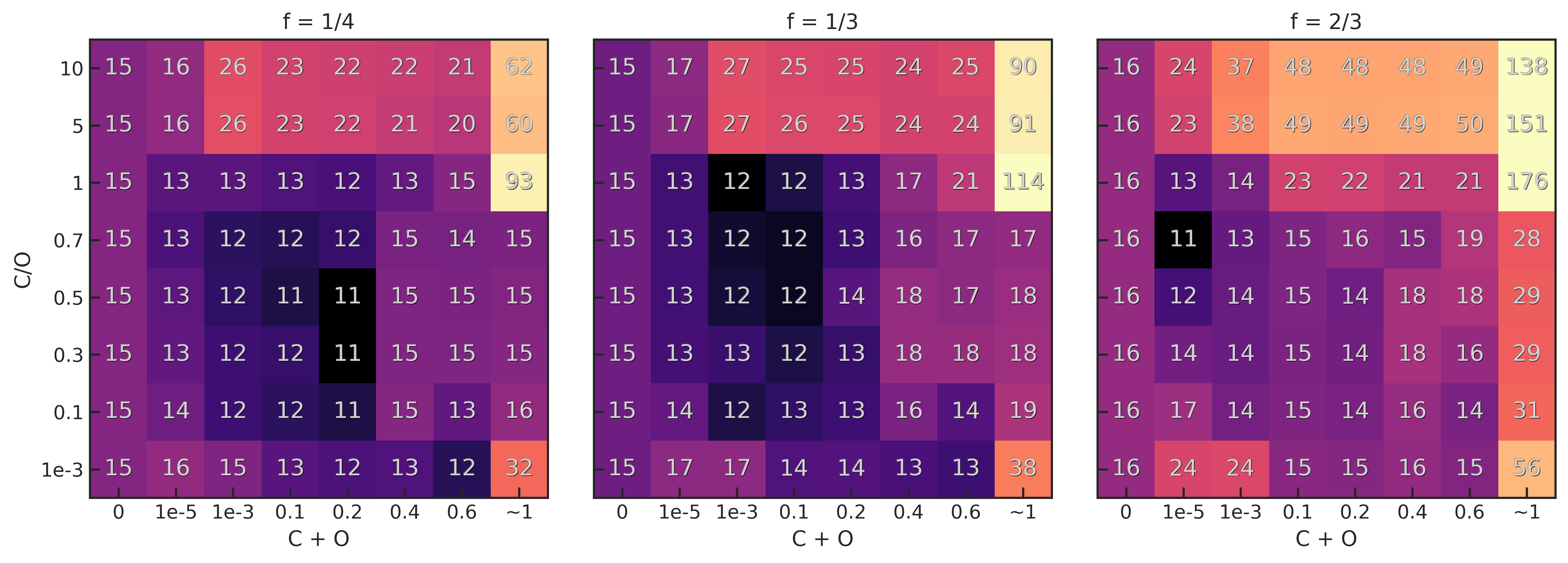}
    \caption{Distribution of $\chi^2$ values as a function of carbon and oxygen mole fractions and the C/O ratio for the \texttt{SPARTA} reduction of the NIRCam data from \citet{Hu_2024}. Each panel corresponds to a specific temperature regime, indicated by the $f$ value above it. The colour gradient and numbers denote the minimum $\chi^2$ values for each case.}
    \label{fig:A4}
\end{figure*}

In Figures \ref{fig:A3} and \ref{fig:A4}, we show the $\chi^2$ distributions for the \texttt{SPARTA} reduction of the combined NIRCam + MIRI spectra and the NIRCam-only spectra, respectively, based on data from \cite{Hu_2024}.

For the combined NIRCam and MIRI spectrum (Fig. \ref{fig:A3}), the trends are broadly consistent with those observed in the \texttt{Eureka!} reduction. The best-fitting cases for $f=1/4$ are centred around C+O mole fractions of 0.1--0.2 and a C/O ratio of 0.1--0.7. By comparison, the \texttt{Eureka!} reduction achieves more confined constraints, centring at C+O = 0.1 and C/O = 0.3--0.5. As the temperature increases, discrepancies between the two reductions become apparent. Unlike \texttt{Eureka!}, \texttt{SPARTA} shows a stronger preference for \ce{CO}-rich atmospheres over \ce{CO2}-rich ones, though \ce{CO2}-rich scenarios still achieve good fits. Additionally, \texttt{SPARTA} does not exhibit significant support for hydrocarbon-rich atmospheres with C/O > 1.0 or for models devoid of carbon and oxygen. For the dayside-confined temperature regime ($f=2/3$), the preferred \texttt{SPARTA} models feature C+O mole fractions of $\approx10^{-3}$, fitting the spectral features using a combination of \ce{PH3} and \ce{CO} opacities.

When analysing only the NIRCam spectrum from the \texttt{SPARTA} reduction (Fig. \ref{fig:A4}), we find that the composition is generally poorly constrained, similar to the results for the \texttt{Eureka!} reduction. At $f=1/4$, the preferences between the two reductions are somewhat aligned, favouring atmospheres containing \ce{CO} and \ce{CO2}. However, \texttt{SPARTA} exhibits a stronger preference for \ce{H2}-containing atmospheres, resulting in a much broader range of viable atmospheric compositions (as discussed in Section \ref{section:alt_reductions}). As the temperature increases to $f=1/3$, the degeneracy becomes more pronounced, with the lowest $\chi^2$ values corresponding to \ce{CO}-, \ce{PO}-, and \ce{PH3}-rich scenarios, offering essentially no constraints on the C+O mole fraction. At $f=2/3$, substantial differences arise between the \texttt{Eureka!} and \texttt{SPARTA} reductions. \texttt{SPARTA} shows more preference for \ce{CO}-dominated atmospheres, which have $C+O = 10^{-5}$ mole fraction and a C/O ratio of 0.7.

\subsection{C+O and C/O trends for the \texttt{stark} reduction}

\begin{figure*}[h!]
    \centering
        \includegraphics[width=1\textwidth]{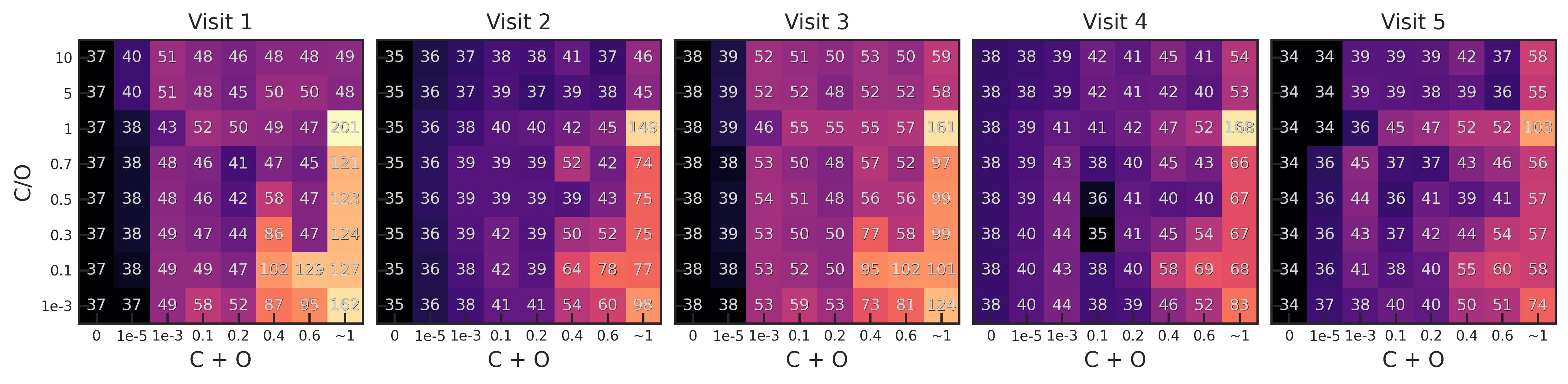}
    \caption{Achieved $\chi^2$ values for varying carbon and oxygen mole fractions as a function of the C/O ratio for the \texttt{stark} reduction of the five NIRCam visits from \citet{Patel_2024}, combined with the MIRI spectrum from \citet{Hu_2024}. Results are shown for $f=1/4$. The colour gradient and numerical labels indicate the minimum $\chi^2$ values for each case.}
    \label{fig:A5}
\end{figure*}

\begin{figure*}[h!]
    \centering
        \includegraphics[width=1\textwidth]{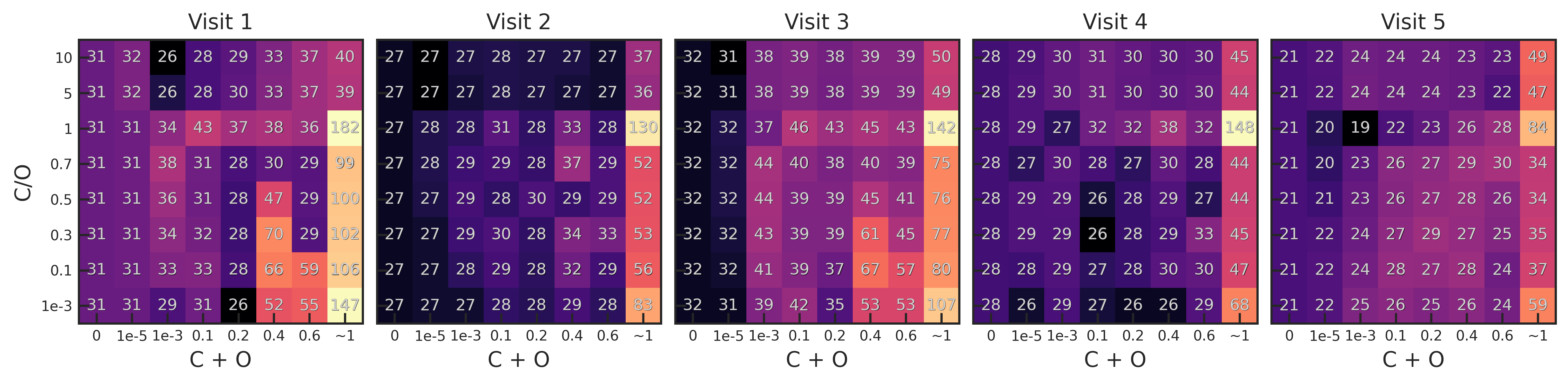}
    \caption{Achieved $\chi^2$ values for the \texttt{stark} reduction of the five NIRCam visits from \citet{Patel_2024}, as in Figure \ref{fig:A5}, but excluding the MIRI spectrum.}
    \label{fig:A6}
\end{figure*}

Figures \ref{fig:A5} and \ref{fig:A6} present a similar analysis for the \texttt{stark} reduction of the five NIRCam visits from \citet{Patel_2024}, focusing on trends at $f=1/4$. Each visit exhibits a distinct spectral shape, leading to varying compositional preferences. Notably, the isolated NIRCam spectra for all visits, except Visit 4, show a tendency towards decreasing C+O abundances or the absence of carbon and oxygen entirely. For Visit 1, the best-fitting scenarios are partially inverted \ce{PH3}-rich atmospheres, followed by inverted \ce{CO} cases with $C+O = 0.2$. Similar partially inverted scenarios are preferred for Visits 2, 3, and 5. Visits 2 and 5 additionally favour models with $C+O \leq 10^{-5}$, where \ce{CO} becomes a dominant absorber. Visit 4 aligns with the data from \citet{Hu_2024} and is best explained by a hydrogen-free, \ce{CO2}- and \ce{PO}-rich atmosphere, consistent with Fig. \ref{fig:A1}.

The individual NIRCam preferences for the \texttt{stark} reduction is shown in Fig. \ref{fig:A6}. For Visit 1, there is a preference for compositions with C/O > 1, corresponding to \ce{HCN}-rich atmospheres (Fig. \ref{fig:F8}), or alternatively, inverted \ce{CO} cases with $C+O = 0.2$. Visit 2 is well-described by a simple blackbody model, showing poor constraints on carbon and oxygen chemistry. However, the best fits include either \ce{CO}- and \ce{HCN}-containing compositions with $C+O \leq 10^{-5}$ or C+O-dominated atmospheres with comparable $\chi^2$ values. Visit 3 trends closely resemble those of the combined NIRCam and MIRI spectrum, with the inverted feature explained by \ce{PH3} or \ce{N2-N2} opacity. For Visit 4, \texttt{stark} and \texttt{Eureka!} reductions show consistent carbon and oxygen trends, differing only by overall increased $\chi^2$ values for the \texttt{stark} reduction. Finally, Visit 5 is best fitted by \ce{CO}- and \ce{PH3}-rich atmospheres, though other compositions show similar $\chi^2$ values.

Different heat redistribution values lead to variations in compositional preferences, with some cases even showing lower $\chi^2$ values. Nevertheless, the overall trends remain consistent with those described for other reductions of 55 Cancri e data, which also incorporate varying heat redistribution values.

\end{appendix}
\end{document}